\documentclass[12pt,preprint]{aastex}
\usepackage{natbib}

\slugcomment{Draft: \today}
\shorttitle{SDSS EDR Quasar Equivalent Widths}
\shortauthors{Wiegert}

\newcommand{\mgii}{Mg~\textsc{ii}}
\newcommand{\ciii}{C~\textsc{iii}]}
\newcommand{\civ}{C~\textsc{iv}}
\newcommand{\oiii}{[O~\textsc{iii}]}
\newcommand{\feii}{Fe~\textsc{ii}}
\newcommand{\uband}{{u^\ast}}
\newcommand{\gband}{{g^\ast}}
\newcommand{\rband}{{r^\ast}}
\newcommand{\iband}{{i^\ast}}
\newcommand{\zband}{{z^\ast}}
\newcommand{\Msolar}{M_\sun}

\newcommand{\mean}[1]{\left\langle{#1}\right\rangle}
\newcommand{\sgn}{\mbox{sgn}}

\begin{document}

\title{Constraining Compact Dark Matter with Quasar Equivalent Widths from
  the Sloan Digital Sky Survey Early Data Release}
\author{Craig C. Wiegert}
\affil{Department of Physics, Enrico Fermi Institute}
\affil{The University of Chicago, Chicago, IL 60637}
\email{wiegert@oddjob.uchicago.edu}

\begin{abstract}
  Cosmologically distributed compact dark matter objects with masses in
  the approximate range of $0.001\Msolar$ to $1\Msolar$ can amplify the
  continuum emission of a quasar through gravitational microlensing,
  without appreciably affecting its broad emission lines.  This will
  produce a statistical excess of weak-lined quasars in the observed
  distribution of spectral line equivalent widths, an effect that scales
  with the amount of compact dark matter in the universe.  Using the
  large flux-limited sample of quasar spectra from the Early Data
  Release of the Sloan Digital Sky Survey (SDSS), I demonstrate the
  absence of a strong microlensing signal.  This leads to a constraint
  on the cosmological density of compact objects of $\Omega_c < 0.03$,
  relative to the critical density, on Jupiter- to solar-mass scales.
  For compact objects clustered in galaxy halos, this limit is expected
  to be weaker by at most a factor of two.  I also forecast the
  improvements to this constraint that may be possible in a few years
  with the full SDSS quasar catalog.
\end{abstract}



\section{Introduction}
\label{cha:intro}

Decades after dark matter's existence was first inferred, its nature and
composition remain largely a mystery.  In particular, the question of
how much of the dark matter is in microscopic vs.\ macroscopic compact
(MACHO) form is still unsettled.  Whatever the form, most of the dark
matter must be non-baryonic.  Measurements of primordial deuterium
abundance in the context of standard Big Bang nucleosynthesis
\citep*{misc:omegaB-BBN} and recent observations of the cosmic microwave
background by WMAP \citep{misc:wmap} limit the baryon density to
$\Omega_B h^2 = 0.02$ in a universe with an overall mass density of
$\Omega_M = 0.3$.  Moreover, the most recent local ``baryon census''
accounts for about 40\% of these baryons in the form of luminous stars,
hot cluster gas, and the Lyman-alpha forest, with as much as 40\%
additionally present in a predicted warm/hot intergalactic medium
(\citealt{misc:silk-baryons} and references therein).

These results do not however rule out compact dark matter objects.  The
baryon census shortfall of approximately 20\% leaves open the
possibility of conventional baryonic MACHO candidates such as brown
dwarfs, Jupiters, low-mass stars, and stellar remnant black holes.
Although baryonic MACHOs would constitute little of the overall dark
matter density, non-baryonic compact dark matter such as primordial
black holes \citep{misc:pbh} or quark nuggets \citep{misc:quark-nuggets}
could substantially contribute.

In this paper I examine the possibility that quasars are
gravitationally microlensed by a cosmological population of compact
objects in the $0.001\Msolar$--$1\Msolar$ range, and that this effect is
statistically detectable in a flux-limited sample of quasar spectra.
Microlensing can be considered a variant of strong lensing, in which the
gravitational deflection due to a compact object produces multiple
images of a background source that are too close to be resolved,
yielding effectively a single magnified image.  Properties of source
distributions may be modified by this magnification in statistically
detectable ways.  

In addition to the statistical effects that will be considered here,
microlensing also generally produces time-dependent magnifications,
typically on scales of days to years, as lenses move in and out of lines
of sight.  This variability is the key to the searches for microlensing
of sources in the Large Magellanic Cloud
\citep{lens:macho-lmc2000,lens:eros-lmc2000} and the galactic bulge
\citep{lens:macho-bulge2000}, due to compact dark objects in our
Galaxy's halo and/or disk.  The results rule out planetary-mass
($\lesssim 0.001\Msolar$) compact objects as a significant component of
the halo but suggest that $0.5\Msolar$ objects could constitute
one-third of the halo population, although this interpretation is still
a subject of debate and does not necessarily imply constraints on
cosmological populations.  Characteristic lightcurves of microlensing
have also been observed within one of the multiple images of strongly
lensed quasars \citep{qso:ogle-qso2237}.  Additionally,
\citet{lens:qso-var-redux} demonstrate the constraints that are possible
from a search for microlensing-induced variability in the larger quasar
population using a method discussed by \citet{lens:qso-var}.  Finally,
on much shorter timescales of seconds to hours, the absence of
time-dependent microlensing in gamma-ray burst observations places weak
constraints on compact objects in several mass ranges from
$10^{-16}\Msolar$ to $10^8\Msolar$ \citep{lens:grb}.

The statistical microlensing method for quasar sources, as developed in
\citet{lens:canizares} and \citet{lens:dalcanton} and employed here,
starts with the fact that the angular scale for lensing by a point mass
(the so-called Einstein angle $\alpha_E$) depends on the mass of the
lens.  This corresponds to a physical size scale in the source plane of
roughly $\eta_{lens} \sim 0.025 h^{-1/2} (M/\Msolar)^{1/2}$~pc for a
compact lens of mass $M$.  Light from regions closer to the optical axis
than this will be strongly magnified, while emission from outside this
area will remain relatively unaffected.  For an extended source of
angular size $\beta_S$ and physical size $\eta_S$, the maximum
magnification of this central region in the source plane is $\mu_{max}
\sim 2\alpha_E/\beta_S = 2\eta_{lens}/\eta_S$.

Quasars are thought to have a small central continuum emitting region
(CR) powered by a supermassive black hole, and surrounded by a much
larger broad-line emission region (BLR).  If $\eta_{CR} \lesssim
\eta_{lens} \lesssim \eta_{BLR}$, a lens can magnify the continuum
background of a quasar spectrum while leaving the emission lines
virtually untouched.  In other words, the lens scale (or equivalently
mass) needs to be large enough to have a significant effect on the CR,
without being so large that it also affects the BLR, for it to be
possible to detect lensing in the spectrum of a quasar.  Variability
timescales of the continuum, in addition to observations of stellar
microlensing in strongly lensed quasars, indicate that the optical
continuum emitting region is smaller than $3\times 10^{-4}$~pc
\citep{qso:cr-shalyapin}.  Line variability times and the correlation
times between continuum and line fluctuations suggest a scale for the
BLR of about $0.1$~pc for quasars at moderate redshift
\citep{qso:blr-gondhalekar}.  Additionally, there is a much larger
narrow-line emission region (NLR) \citep{book:agn-peterson} that may
contribute to the emission line flux.  These length scales imply that
microlensing can occur for lens masses within a few decades around
$\Msolar$, encompassing a wide variety of compact dark matter
candidates.

Thus, while gravitational deflection of light is an achromatic
phenomenon, microlensing by compact objects can still alter the spectrum
of a distant quasar.  The statistical spectral signature of lensing will
be the presence of weak-lined (small equivalent width) objects in a
flux-limited sample of quasars.  This signal increases both for higher
source redshift and with larger compact dark matter fraction $\Omega_c$.
Because of this, knowing the fraction of small equivalent width quasars
over a wide range of redshifts can place constraints on $\Omega_c$.

Previous efforts using this method to obtain limits on compact dark
matter from quasar observations have produced upper bounds of $\Omega_c
< 0.2$ for compact lenses with masses from $0.001$--$60\Msolar$
\citep{lens:dalcanton}, but suffer from small-number statistics ($\sim
200$~quasar lines).  With the orders-of-magnitude larger quasar survey
results now available or in the works, a more accurate quasar optical
luminosity function, and a better understanding of the likely
cosmological model, the subject deserves to be revisited.  I apply the
technique to the quasar catalog from the Early Data Release (EDR) of the
Sloan Digital Sky Survey (SDSS).  In addition to deriving stronger
constraints on $\Omega_c$ using this sample, I forecast the limits
possible from the upcoming deluge of SDSS spectroscopic data.

Section~\ref{cha:theory} develops the theory needed to model the
distribution of quasar equivalent widths in the presence of microlenses.
I present the details of the EDR quasar sample in
Section~\ref{cha:data}, and discuss its strengths and limitations with
respect to the lensing model.  In Section~\ref{cha:analysis} I apply the
model to the data, and Section~\ref{cha:conclude} summarizes the results
and suggests possible systematic effects to be considered in future
work using this method.

\section{Theoretical Model}
\label{cha:theory}

The goal of this section is to calculate a theoretical expression for
the observed distribution of quasars equivalent widths, $p_W(W) \,dW$,
given some unlensed intrinsic equivalent width distribution $p_{W0}(W_0)
\,dW_0$.  This derivation follows the ones found in \cite{lens:dalcanton}
and \citet{lens:canizares} and includes the effects of amplification
bias and extended sources.  Throughout the discussion, unless
specifically noted otherwise, I assume a currently favored flat
lambda-dominated model in which $\Omega_\Lambda = 0.7$ and $\Omega_M =
0.3$, rather than the Einstein-de Sitter models found in the
previous papers.


\subsection{Microlensing Probability}
\label{sec:mulens}

The first task is to compute the probability $p(\mu;z) \,d\mu$ that an
object at redshift $z$ is magnified by a factor $\mu$.  An analytically
simple lensing model is chosen, in which point-mass (Schwarzschild)
lenses are assumed to be distributed uniformly with constant comoving
density.  The optical depth of the lenses yields the Poisson probability
that a given number of lenses lies along the line of sight.  Then
$p(\mu;z) \,d\mu$ is the sum of the probability for magnification due to
$n$ lenses weighted by the Poisson probabilities.

As is common in derivations of scattering probability, the lensing cross
section (and thus the optical depth) diverges at large impact parameter.
Therefore, define $\xi_0(z'; z)$ to be the maximum impact parameter, in
the lens plane, for a lens at redshift $z'$ and a source at $z$.  Lines
of sight that pass within $\xi_0$ of a lens will be magnified by some
amount greater than $\mu_0$ and are considered to be lensed for purposes
of computing the optical depth; those lines of sight outside $\xi_0$ are
treated as unlensed.  Choose a conveniently small minimum magnification
of $\mu_0 = 1.061$, the magnification at which the lens-quasar intrinsic
angular separation is equal to a critical angle $\alpha_0$ that is twice
the Einstein angle:
\begin{equation}
  \label{eq:crit-angle}
  \alpha_0^2 = (2\alpha_E)^2 = \frac{16 GM}{c^2}\frac{D_{LS}}{D_L D_S}.
\end{equation}
Here $M$ is the lens mass and $D$ is the angular diameter distance,
where the subscripts $L$, $S$, and $LS$ refer to observer-to-lens,
observer-to-source, and lens-to-source distances respectively.  Note
that $\xi_0 = \alpha_0 D_L$.  The details of the choice of these
low-magnification parameters $\mu_0$ and $\alpha_0$ will not
substantially affect the lensed equivalent width distribution in the
end.

There is some uncertainty regarding the particular choice of angular
diameter distance in Equation~\ref{eq:crit-angle} and in general for
lensing calculations.  It is argued that since lensing necessarily
represents a departure from a perfectly homogeneous
Friedmann-Lema\^itre-Robertson-Walker (FLRW) universe, one must use for
example the Dyer-Roeder ``empty cone'' distances
\citep{lens:dyer-roeder}.  However, \citet{book:peacock} makes the
argument that for the vast majority of lensing calculations, it is most
appropriate simply to use FLRW distances; that is the choice I have made
here.  The effect on the lensing calculations is typically only on the
order of a few percent, and in an already simplified lensing model, the
choice of angular diameter distance will not have a large impact on the
final result.  The FLRW angular diameter distance along a geodesic
between $z_1$ and $z_2$ can be written as
\begin{equation}
  \label{eq:angdist}
  D(z_1, z_2) = \frac{c}{H_0\,(1+z_2)} \sqrt{\frac{k}{\Omega-1}} \:
  S_k\!\!\left( \sqrt{\frac{\Omega-1}{k}} 
    \int_{z_1}^{z_2} \frac{H_0 \,dz}{H(z)} \right) ,
\end{equation}
where $k =\sgn(\Omega-1)$ is the curvature parameter, $S_k(r) = \{\sinh
r, \,r, \,\sin r\}$ for $k = \{-1,\,0,\,1\}$, and
\begin{equation}
  \label{eq:hubble}
  H(z) = H_0 \sqrt{\Omega_M(1+z)^3 + \Omega_\Lambda + (1-\Omega)(1+z)^2}
\end{equation}
is the redshift-dependent Hubble expansion rate in a matter$+\Lambda$
($\Omega = \Omega_M+\Omega_\Lambda$) universe.  When $\Omega = 1$, the
integral in Equation~\ref{eq:angdist} is particularly easy to express in
terms of an elliptic integral of the first kind (see Equation~3.139.2 of
\citealt{book:gr}).
  
The probability $p_u(\mu_u,z';z) \,d\mu_u\,dz'$ of a single lens at $z'$
magnifying an object at $z$ by a factor $\mu_u$ separates into two
components: $p_1(\mu_u)\,d\mu_u$, the probability of one lens causing a
magnification $\mu_u$; and $p_z(z';z)\,dz'$, the probability of the lens
being within range of a line of sight.  The subscript $u$ indicates that
the lensing probabilities have not yet been corrected for flux
conservation.  For a single point-mass lens,
\begin{equation}
  \label{eq:p1}
  p_1(\mu_u) d\mu_u = \cases{
    \frac{\sqrt{\mu_0^2-1}}{\mu_0-\sqrt{\mu_0^2-1}}
    (\mu_u^2-1)^{-3/2} \,d\mu_u 
    & for $\mu_u \geq \mu_0$ \cr
    0
    & otherwise.\cr}
\end{equation}
(see for example \citealt{lens:pei} or \citealt*{book:sef}).
For the chosen minimum magnification of $\mu_0=1.061$, the normalization
prefactor in the above equation is 1.992.
The line-of-sight probability factor, which contains all the redshift
information, is just the product of the number density of lenses and the
lensing cross section:
\begin{equation}
  \label{eq:pz-first}
  p_z(z';z) dz' = n(z') \,\pi \xi_0^2 \,\frac{dr_{prop}}{dz'} \,dz' .
\end{equation}
For a constant comoving lens density ($n(z') = n_0\,(1+z')^3$), and using
the relation between proper distance and redshift $dr_{prop}/dz' =
c/[(1+z')H(z')]$, Equation~\ref{eq:pz-first} can be rewritten as
\begin{equation}
  \label{eq:pz}
  p_z(z';z) dz' = \frac{6 \Omega_c\,H_0^2}{c\,H(z')}\frac{(1+z')^2 D_L
  D_{LS}}{D_S}\,dz' .
\end{equation}
Note that although this derivation assumes that all lenses have the same
mass $M$, the result is unchanged by instead integrating over some mass
distribution function \citep{lens:press-gunn}.

\begin{figure}
  \plotone{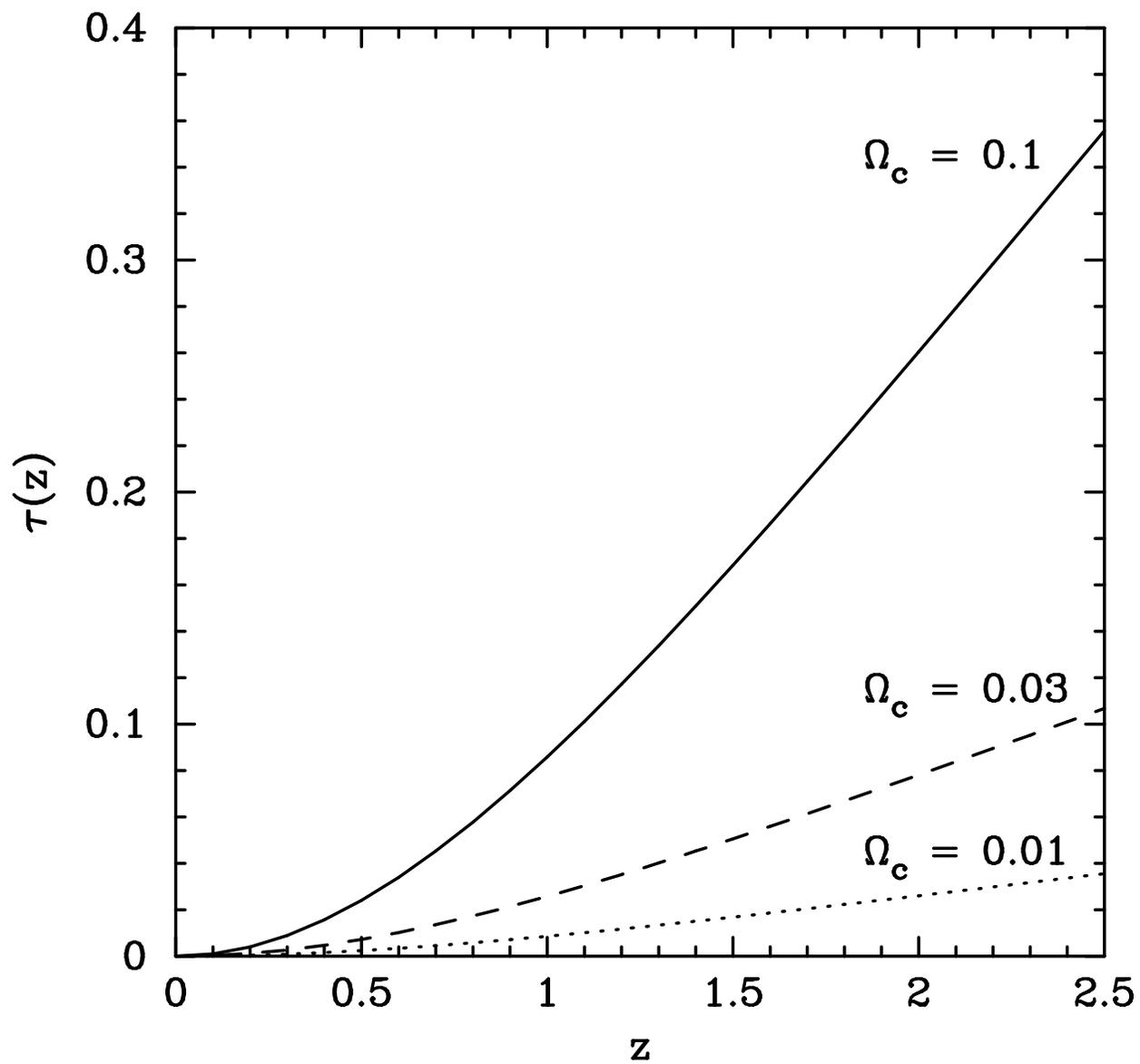}
  \caption{Optical depth as a function of redshift for 
    $\Omega_c = 0.1$ (\emph{solid line}), 
    $\Omega_c = 0.03$ (\emph{dashed line}), and
    $\Omega_c = 0.01$ (\emph{dotted line}).}
  \label{fig:opt-depth}
\end{figure}

Integrating Equation~\ref{eq:pz} over the lens redshift $z'$ yields the
optical depth: 
\begin{equation}
  \label{eq:tau}
  \tau(z) = \int_0^z p_z(z';z) \,dz' .
\end{equation}
Figure~\ref{fig:opt-depth} shows the optical depth plotted for various
values of $\Omega_c$ in the given cosmological model.  From the optical
depth, one can calculate the Poisson probability of a line of sight
encountering a given number of lenses:
\begin{equation}
  \label{eq:poisson}
  P_n(z) = \frac{e^{-\tau(z)} \tau(z)^n}{n!} .
\end{equation}

Now, the probability of a point source at $z$ being magnified by $\mu_u$ can
be written formally as
\begin{equation}
  \label{eq:pu-formal}
  p_u(\mu_u;z) \,d\mu_u = \sum_{n=0}^{\infty} P_n(z) p_n(\mu_u) \,d\mu_u ,
\end{equation}
where $p_n(\mu_u)$ is the probability for $n$ lenses to produce a total
magnification of $\mu_u$.  Given that the optical depth is
small for any realistic choice of $\Omega_c$ at low to moderate redshift
(see Figure~\ref{fig:opt-depth}), multiple-lensing events will be quite
rare, so Equation~\ref{eq:pu-formal} is
well-approximated by
\begin{equation}
  \label{eq:pu-approx}
  p_u(\mu_u;z) = P_0(z)p_0(\mu_u) + P_1(z)p_1(\mu_u) +
  \left[1-P_0(z)-P_1(z)\right] p_{\geq 2}(\mu_u) ,
\end{equation}
where $p_1(\mu_u)$ is given in Equation~\ref{eq:p1},
\begin{equation}
  \label{eq:p0}
  p_0(\mu_u) \,d\mu_u = \delta(\mu_u - 1) \,d\mu_u ,
\end{equation}
and
\begin{equation}
  \label{eq:pg2}
  p_{\geq 2}(\mu_u) \,d\mu_u = \cases{
    2\mu_0^4 \,\mu_u^{-3} \,d\mu_u & for $\mu_u \geq \mu_0^2$\cr
    0 & otherwise.\cr
  }
\end{equation}
Equation~\ref{eq:pg2} is an approximate expression from
\citet{lens:canizares} for the probability of magnification from
multiple lenses, found by convolving two single lens probabilities and
assuming that one of the individual magnifications is weak.  As
discussed in \citet{lens:pei}, this is a good assumption representing
the vast majority of multiple-lens events.

The derivation so far has neglected the matter of flux conservation.
Because gravitational lensing conserves photon number and energy
\citep{lens:flux-cons}, it is easy to see that the magnification must
satisfy $\mean{\mu} = 1$ when averaged over all lines of sight.  Flux is
not explicitly conserved in the expressions above because only
overdensities (i.e., the compact lenses) were taken into account, and
not the counterbalancing underdense regions.  An approximate solution to
this concern, following \citet{lens:canizares}, is to compensate by
scaling the magnification by a redshift-dependent diminution factor
$\mu_z$, defined so that
\begin{equation}
  \label{eq:mu-z}
  \frac{1}{\mu_z} = \mean{\mu_u} =
  \int_0^\infty \mu_u \, p_u(\mu_u; z) \,d\mu_u .
\end{equation}
Using this correction and the substitution $\mu = \mu_z \mu_u$,
\begin{eqnarray}
  \label{eq:p-final}
  p(\mu;z) \,d\mu & = & p_u(\mu_u;z) \,d\mu_u \nonumber\\
  & = &
  \frac{1}{\mu_z} \left[
    P_0(z)p_0\!\left(\frac{\mu}{\mu_z}\right) + 
    P_1(z)p_1\!\left(\frac{\mu}{\mu_z}\right) + \right. \\
  & &  \left. 
  \left(1-P_0(z)-P_1(z)\right)p_{\geq 2}\!\left(\frac{\mu}{\mu_z}\right)
  \right] d\mu . \nonumber
\end{eqnarray}
The correction is not large; the upper limit for present purposes is
$\mean{\mu_u} = 1.17$, calculated for $z=2.5$ and assuming $\Omega_c=0.1$.
\citet{lens:dalcanton} argue in detail that this approximation is both
self-consistent and appropriate when considering even moderate
magnifications of $\mu > 1.5$.  It is also worth pointing out that the
much larger scale of underdense regions relative to the lenses implies
that this approximate treatment is as valid for the BLR as for the CR.
In the low-magnification regime, which is not of interest here, there
are many additional uncertainties and approximations (e.g.,
inhomogeneous lens distributions, beam shear, source surface brightness,
etc.).

\subsection{Extended Sources}
\label{sec:ext-source}

For some applications the point source assumption in
Section~\ref{sec:mulens} is valid.  However, as discussed in
Section~\ref{cha:intro}, the spectral signature of compact dark matter
occurs when the physical scale of lensing is bounded above and below by
the sizes of the quasar broad-line emission and continuum regions
respectively.  Thus it is necessary to modify the above equations to
account for the reality of an extended source.

\citet{lens:schneider-ext-source} shows that the magnification of a
uniformly bright circular source decreases monotonically as the distance
of the source from the optical axis increases.  Because of this, the
magnification probability rapidly decays to zero at some maximum
magnification, occurring when the source is collinear with the lens and
observer:
\begin{equation}
  \label{eq:magmax}
  \mu_{max} = \sqrt{1+\left(\frac{2\alpha_E D_S}{\eta_S}\right)^2} .
\end{equation}
Here $\eta_S$ is the radius of the uniform source.
Equation~\ref{eq:magmax} can be rewritten as
\begin{equation}
  \label{eq:magmax2}
  \mu_{max} = \sqrt{1+\left(\frac{\xi_{0,proj}}{\eta_S}\right)^2} ,
\end{equation}
where $\xi_{0,proj} = \xi_0 D_S/D_L$ is the projection of the maximum
impact parameter from the lens plane to the source plane.  

\begin{figure}
  \plotone{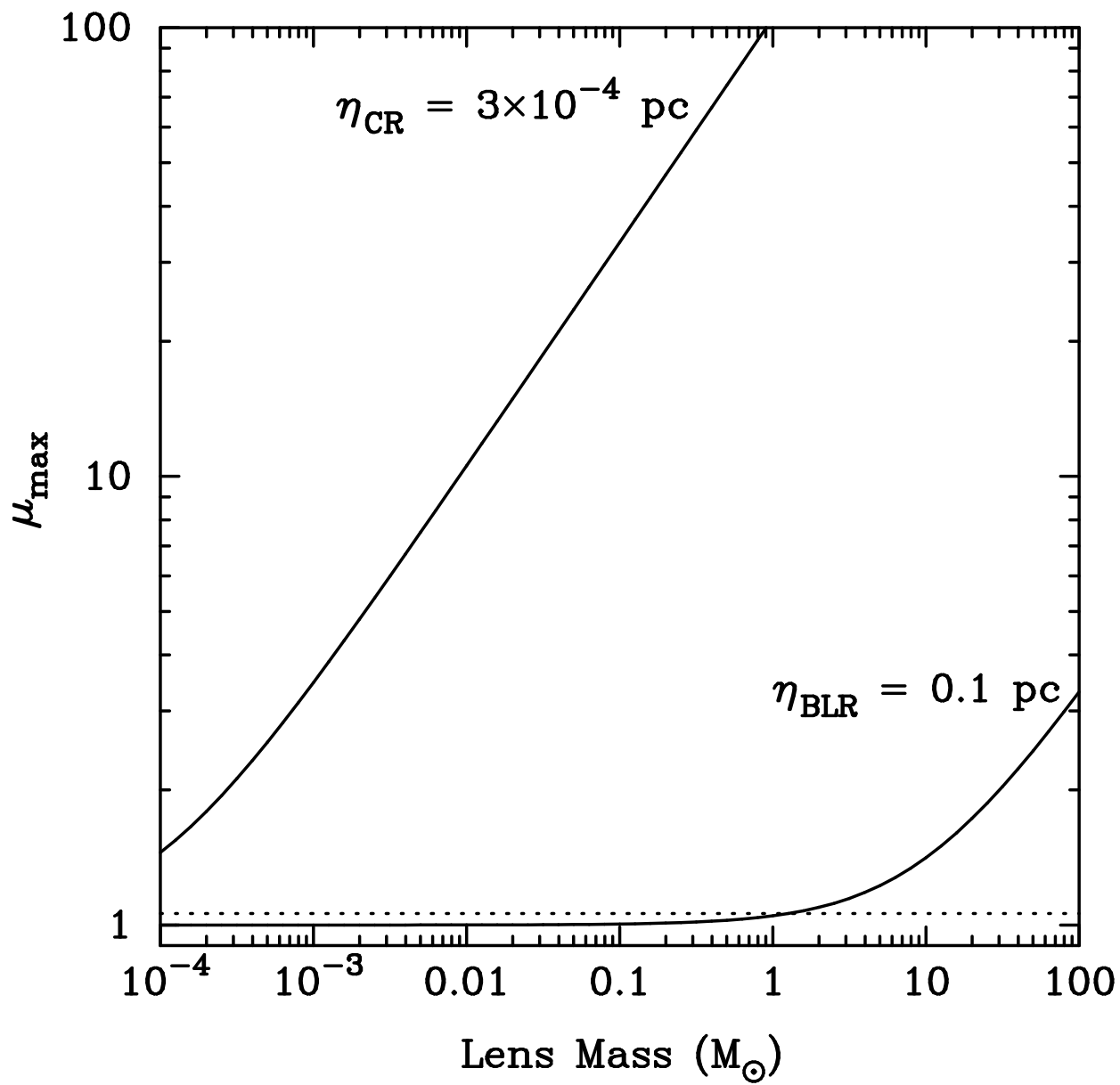}
  \caption{Average maximum magnification as a function of lens mass, for
    different choices of source size, plotted for $z=2$.  The dotted
    horizontal line indicates the minimum magnification of the lensing
    model, $\mu_0 = 1.061$.}
  \label{fig:mean-magmax}
\end{figure}

As written, $\mu_{max}$ is a function of both the lens redshift $z'$ and
the source redshift $z$, as well as the lens mass $M$ and source radius
$\eta_S$.  In order to incorporate this magnification limit into the
discussion of lensing statistics, it is first necessary to average over
$z'$, weighting by the lens probability $p_z(z';z)$.  However, averaging
$\mu_{max}$ directly is both unwieldy and unenlightening, so instead
compute the average of $\xi_{0,proj}$:
\begin{equation}
  \label{eq:l0proj}
  \mean{\xi_{0,proj}} = \frac{6\Omega_c \,H_0^2}{\tau(z) \,c}
  \left(\frac{GM}{c^2} \right)^{1/2}
  \int_0^z \frac{(1+z')^2}{H(z')} 
  \frac{D_{LS}^{3/2}D_L^{1/2}}{D_S^{1/2}} \,dz' .
\end{equation}
This can be substituted into Equation~\ref{eq:magmax2} to obtain an
expression for $\mean{\mu_{max}}$ that, while not strictly speaking the
actual mean, is an extremely close approximation.  Because of the
optical-depth normalization factor in Equation~\ref{eq:l0proj},
$\mean{\xi_{0,proj}}$ is independent of $\Omega_c$ and only weakly
dependent on $z$.  Figure~\ref{fig:mean-magmax} shows how the maximum
magnification depends on the remaining parameters of lens mass and
source radius.

As in \citet{lens:dalcanton}, the rapid decay of probability
distributions $p_1(\mu)$ and $p_{\geq 2}(\mu)$ can be approximated by
applying a cutoff at $\mu_u = \mean{\mu_{max}}$ and adding a normalizing
$\delta$-function at the cutoff.  Given the multiplicative magnitude
assumption in the multiple-lens case, it might make more sense to cutoff
$p_{\geq 2}(\mu)$ above $\mu_{max}^2$, but once again this detail is
irrelevant in light of the small optical depth.


Figure~\ref{fig:mean-magmax} also demonstrates that the size of the
quasar continuum and emission line regions determine the range of
microlens masses to which this lensing test applies.  For very small
masses, the continuum will never be magnified enough to result in
weak-lined quasars.  At large masses, although the maximum continuum
magnification is no longer a concern, the broad line region will also be
substantially magnified.  Based on Figure~\ref{fig:mean-magmax}, I adopt
$1\Msolar$ as a conservative upper bound on lens mass; above this
mass, the broad line region can be magnified above the low-magnification
threshold $\mu_0$ used in Equation~\ref{eq:p1}.  A similarly
conservative lower bound of $0.001\Msolar$ reflects the fact that below
this lens mass, the continuum is almost never magnified enough to
produce a significant signal of weak-lined objects.

\subsection{Amplification Bias}
\label{sec:amp-bias}

In a flux-limited survey, objects that are not intrinsically bright
enough to be observed are sometimes magnified enough by lensing to be
detectable.  Therefore if there is any appreciable amount of lensing, a
flux-limited sample will contain a greater fraction of lensed objects
than might ordinarily be expected.  In this case, the observed
distribution of magnifications in the sample $p_s(\mu;z)$ will be
greater than $p(\mu;z)$ for large $\mu$.  Specifically, if the
luminosity lower limit of the survey is $L_s$ for a given redshift, then
quasars magnified by $\mu$ with a luminosity greater than $L_s/\mu$ will
appear in the sample.  Thus, for a differential luminosity function
$\phi(L,z) \,dL$,
\begin{equation}
  \label{eq:ps}
  p_s(\mu;z) \propto p(\mu;z) \int_{L_s/\mu}^\infty \phi(L,z) \,dL .
\end{equation}
Note that luminosity functions rising steeply with decreasing luminosity
will result in a much larger amplification bias than those with a
shallower slope.

Because the SDSS Collaboration has not yet determined a complete quasar
luminosity function from the Survey data (although a preliminary
determination for high-redshift quasars can be found in
\citet{sdss:edr-hizqso-lf}), I use the optical luminosity function from
the 2dF Quasar Redshift Survey \citep{twodf:lf}.  They find that the
differential luminosity function is best fit by a double power-law
function of the form
\begin{equation}
  \label{eq:qsolf}
  \phi(L_B,z) = 
  \frac{\phi(L_B^\ast)}{(L_B/L_B^\ast)^{\beta_1}+(L_B/L_B^\ast)^{\beta_2}} ,
\end{equation}
or, expressed in absolute $B$ magnitudes, and correcting a misprint in
\citet{twodf:lf}, 
\begin{equation}
  \label{eq:qsolf2}
  \phi(M_B,z) = 
  \frac{\phi(M_B^\ast)}
  {10^{0.4(\beta_1-1)(M_B^\ast-M_B)} +
  10^{0.4(\beta_2-1)(M_B^\ast-M_B)}} .
\end{equation}
The luminosity function is assumed to evolve with redshift, with the
characteristic luminosity at the power-law break given by
\begin{equation}
  \label{eq:LBstar}
  L_B^\ast(z) = L_B^\ast(0)10^{k_1 z + k_2 z^2} ,
\end{equation}
or equivalently
\begin{equation}
  \label{eq:MBstar}
  M_B^\ast(z) = M_B^\ast(0)-2.5(k_1 z + k_2 z^2) .
\end{equation}
For a universe with $\Omega_M = 0.3$, $\Omega_\Lambda = 0.7$, and $H_0 =
50$~km~s$^{-1}$~Mpc$^{-1}$, \citet{twodf:lf} find the best-fit
parameters to be $\beta_1 = 3.41$, $\beta_2 = 1.58$, $k_1 = 1.36$, $k_2 =
-0.27$, $M_B^\ast = -22.65$, and $\phi^\ast = 0.36\times
10^{-6}$~Mpc$^{-3}$~mag$^{-1}$.

With the above functional form of the luminosity function, the observed
magnification distribution can be written in terms of a hypergeometric
function:
\begin{equation}
  \label{eq:ps2}
  p_s(\mu;z) \propto p(\mu;z) 
  \,{}_2F_1 \!\left( 1 , \frac{\beta_1-1}{\beta_1-\beta_2};  
    1 + \frac{\beta_1-1}{\beta_1-\beta_2}; 
    -\!\left(\frac{\mu}{R(z)}\right)^{\beta_1-\beta_2} \right)
  \left(\frac{\mu}{R(z)}\right)^{\beta_1-1}
\end{equation}
where I have defined $R(z)$ to be the ratio of the sample flux limit to
the characteristic luminosity of the luminosity function, $R(z) =
L_s(z)/L_B^\ast(z)$.  In practice, the normalization constant for this
equation can be easily computed numerically by integrating
$p_s(\mu;z)\,d\mu$.


Of course, the observed luminosity function itself will be slightly
modified from the intrinsic form because of lensing.  If $\Omega_c$ is
large enough to cause significant lensing, then the intrinsic luminosity
function will have a steeper slope than what is actually observed
\citep{lens:vietri,lens:schneider-ext-source}.  For the purposes of
calculating the flux ratio $R(z)$ and the amplification bias, it is
assumed that these changes in the luminosity function's slope are of
second order in $\Omega_c$.  Since it is already believed from
\citet{lens:dalcanton} that $\Omega_c < 0.1$, the effect on the
luminosity function would be negligible.  If in fact the intrinsic
luminosity function were steeper, then the amplification bias would be
greater; hence, ignoring the effects of lensing on the luminosity
function leads to more conservative constraints on $\Omega_c$.

\begin{figure}
  \plotone{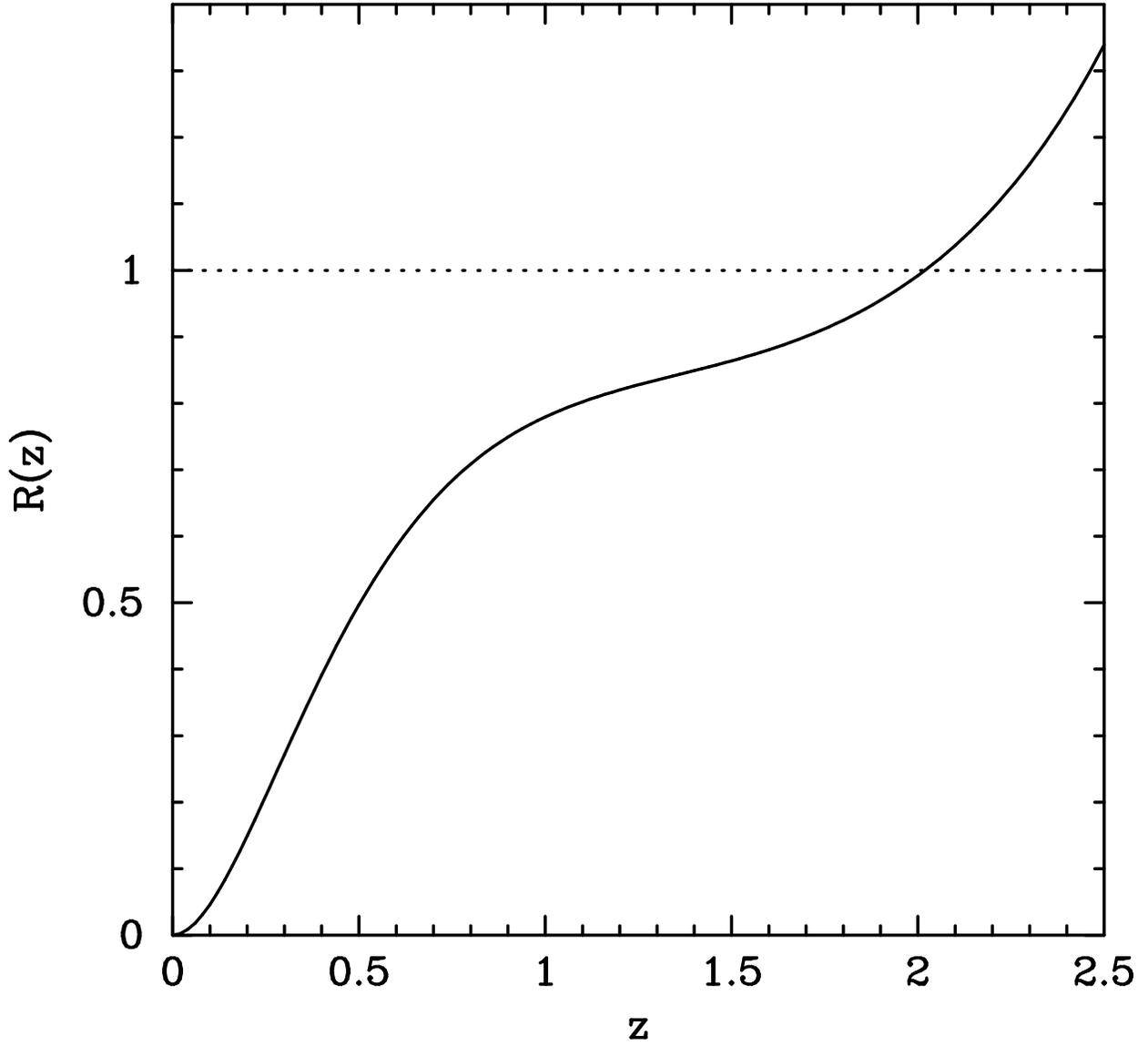}
  \caption{Ratio of the survey flux limit to the characteristic flux of the
    luminosity function, with parameters chosen to match the SDSS EDR
    Quasar Catalog.}
  \label{fig:Rz}
\end{figure}

For a survey with an apparent $B$ magnitude limit $m_s$, the flux ratio
$R(z)$ can be expressed as follows:
\begin{equation}
  \label{eq:Rz}
  -2.5\log R(z) = m_s - M_B^\ast(z) 
  -5\log\left(\frac{d_\ell(z)}{10 \mbox{pc}}\right)
  -2.5(\alpha - 1)\log(1+z) ,
\end{equation}
where $d_\ell(z) = (1+z)^2 D(z)$ is the luminosity distance, and the
last term is the K~correction for an object with spectral energy
distribution $f_\nu \propto \nu^{-\alpha}$, a fairly good approximation
to the quasar UV/optical continuum.  \citet{sdss:qso-composite} find
that $\alpha = 0.5$ for quasars drawn from SDSS data.  As discussed
later in Section~\ref{sec:edr}, the SDSS quasar magnitude limit is
expressed in $\iband$ rather than $B$; I compensate by adding a constant
color term $B-i = 0.35$ to the $\iband$ limit to obtain $m_s$ in the
above equation \citep{sdss:edr-qsocat}.  Figure~\ref{fig:Rz} plots the
flux ratio as a function of redshift for the SDSS Quasar Catalog and
using the 2dF luminosity function.

\subsection{Equivalent Width Distribution}
\label{sec:ew-dist}

The spectral intensity in the vicinity of an emission line is sum of two
components, from the continuum and the emission-line region itself:
\begin{equation}
  \label{eq:spec-intensity}
  I_{tot}(\lambda) = I_{CR}(\lambda) + I_{BLR}(\lambda) .
\end{equation}
The equivalent width of an emission line at wavelength $\lambda_0$ is
defined as the continuum-weighted integral of the line intensity and is
quoted in units of wavelength:
\begin{equation}
  \label{eq:ew-def}
  W = \frac{1}{I_{CR}(\lambda_0)}\int
  \left[ I_{tot}(\lambda)-I_{CR}(\lambda) \right] d\lambda .
\end{equation}
If the continuum region of the quasar is preferentially magnified by a
factor $\mu$, then the observed continuum intensity in the above
equation becomes $\mu \,I_{CR}(\lambda_0)$, while the strength of the line
itself $I_{BLR}(\lambda)$ remains essentially unchanged.  The emission
line of a lensed object is thus effectively demagnified with respect to
its unlensed width $W_0$:
\begin{equation}
  \label{eq:Wlambda}
  W_\lambda = \frac{W_0}{\mu} .
\end{equation}
If $p_{W0}(W_0)\,dW_0$ is the intrinsic distribution of rest-frame
equivalent widths, then the observed rest-frame distribution
$p_W(W_\lambda)\,dW_\lambda$ is:
\begin{equation}
  \label{eq:pW}
  p_W(W_\lambda; z) \,dW_\lambda = dW_\lambda \int_0^\infty \mu 
  \, p_0(\mu W_\lambda) \,p_s(\mu ; z) \,d\mu .
\end{equation}
Equations~\ref{eq:Wlambda} and~\ref{eq:pW} imply that one would expect
an enhanced population of quasars with small equivalent widths if
lensing is significant.

\begin{figure}
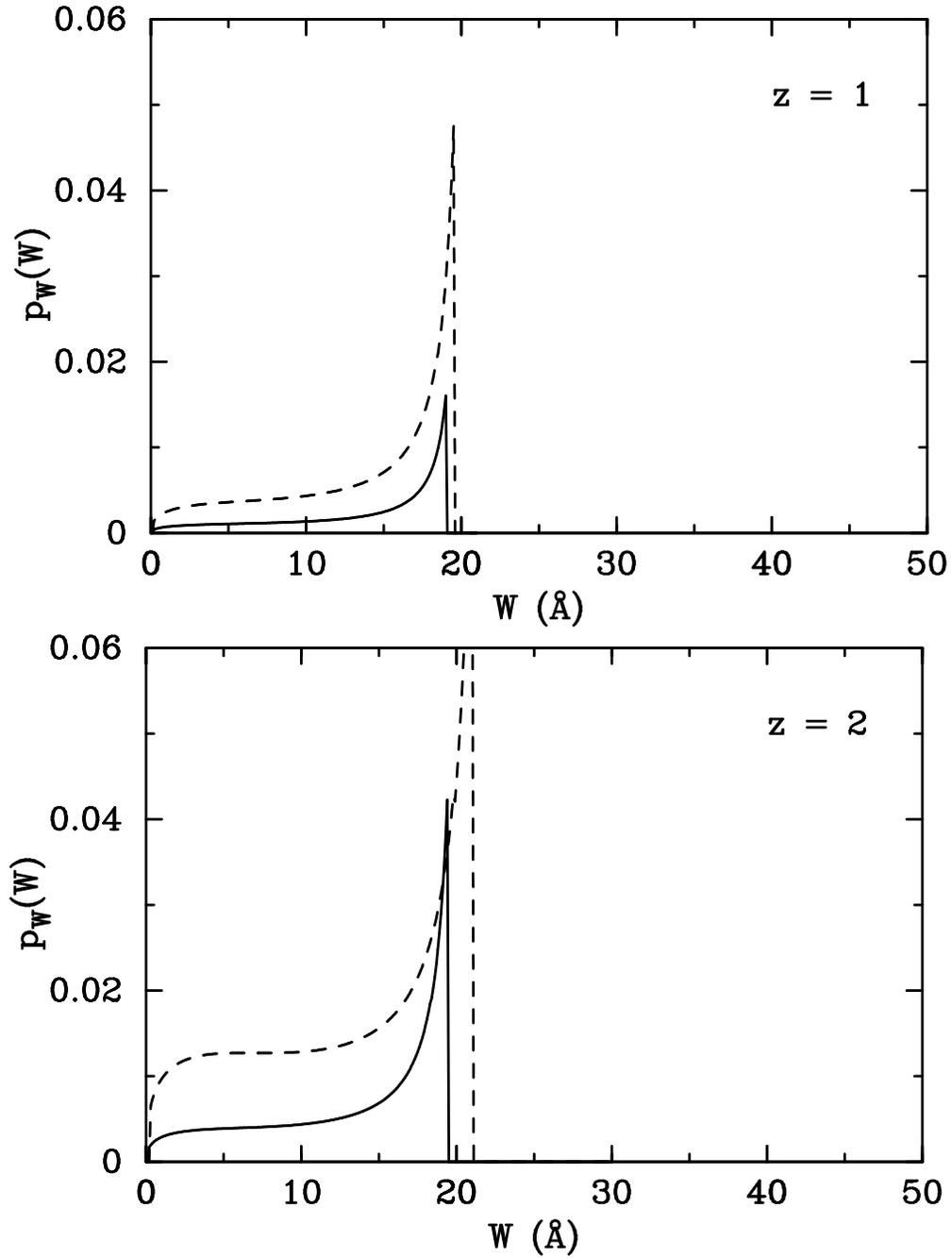

  \centering
  \epsscale{0.8}
  \plotone{f4a.eps}
  \plotone{f4b.eps}
  \epsscale{1.0}
  \caption{Lensed equivalent width distribution, given an intrinsic
    $\delta$-function distribution, for 
    $\Omega_c = 0.03$ (\emph{solid line}) and 
    $\Omega_c = 0.1$ (\emph{dashed line}), 
    at redshifts of 1 (top) and 2 (bottom).  These plots use the SDSS
    EDR magnitude limit for quasars.}
  \label{fig:ew-dist-delta}
\end{figure}

To examine the behavior of Equation~\ref{eq:pW}, consider a highly
unrealistic $\delta$-function intrinsic equivalent width distribution.
The lensing model predictions for a population of emission lines with
$W_0=20$~\AA\ are plotted in Figure~\ref{fig:ew-dist-delta} (the
functions appear not to be normalized, due to the inability to plot the
$\delta$-function at $W = \mean{\mu_u}W_0$ from Equation~\ref{eq:p0}).
The low equivalent width signal shown in these plots is a direct
consequence of the bias-amplified, high-magnification tail from
Equation~\ref{eq:ps2}.

\begin{figure}
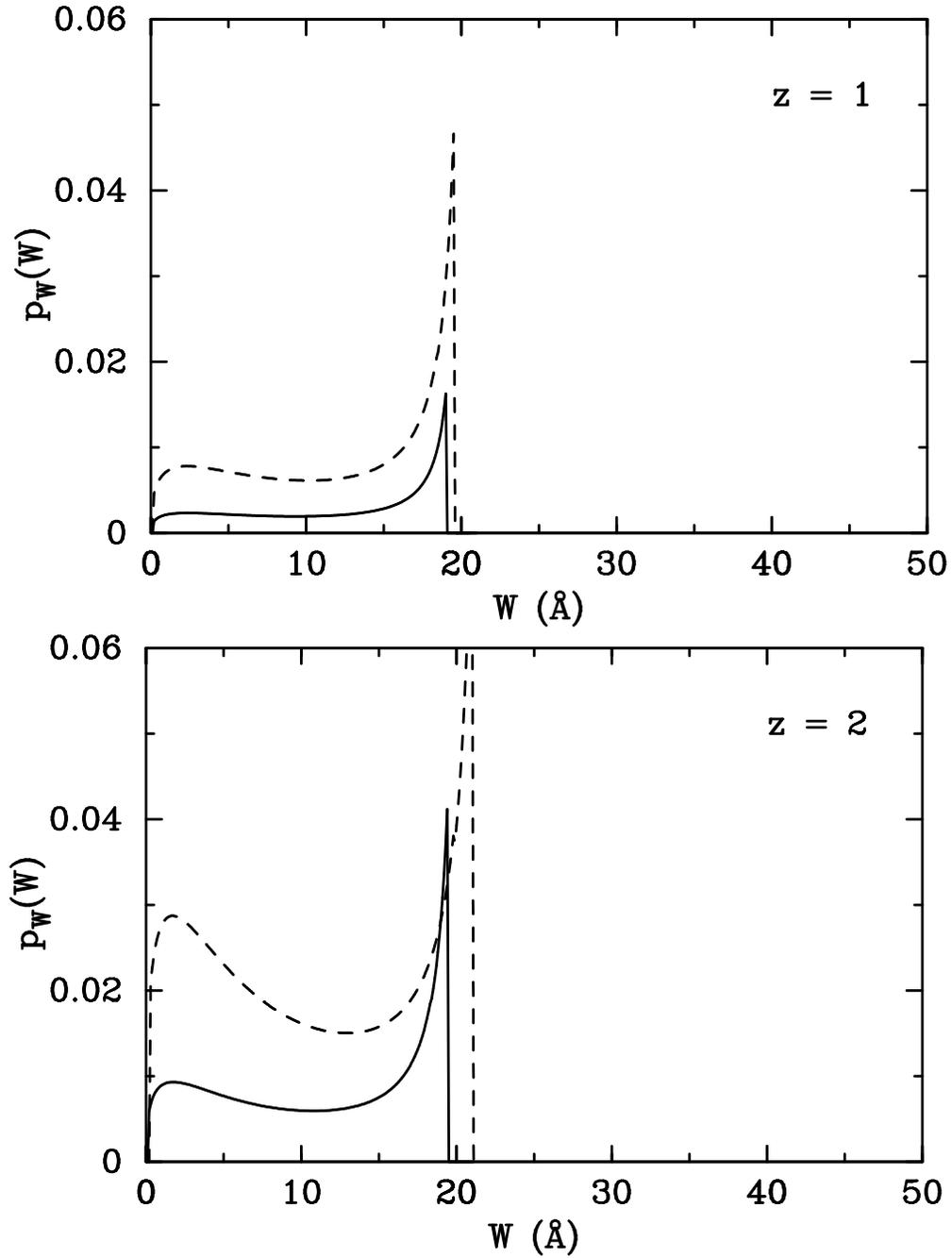

  \centering
  \epsscale{0.8}
  \plotone{f5a.eps}
  \plotone{f5b.eps}
  \epsscale{1.0}
  \caption{Lensed equivalent width distribution, given an intrinsic
  $\delta$-function distribution, for
  $\Omega_c = 0.03$ (\emph{solid line}) and
  $\Omega_c = 0.1$ (\emph{dashed line}), 
  again at redshifts of 1 (top) and 2 (bottom).  Here the survey flux
  limit is set 1~mag brighter to demonstrate the effect of amplification
  bias.}
  \label{fig:ew-dist-delta2}
\end{figure}

It is instructive to consider how the lensed distribution is affected by
the form of the flux ratio $R(z)$.  The value of this function at a
given redshift essentially determines the amount of amplification bias.
For $R(z) > 1$, the survey flux limit is located on a steeper portion of
the luminosity function compared to when $R(z) < 1$, indicating many
more quasars just below the detection threshold that can be lensed into
the sample.  This leads to an interesting conclusion: by setting the
survey flux limit artificially higher, one can expect to see a greater
percentage of microlensed quasars.  A dramatic illustration of this is
shown in Figure~\ref{fig:ew-dist-delta2} for which the survey limit was
made 1~mag brighter relative to Figure~\ref{fig:ew-dist-delta}.  Every
1~mag change in the survey limit multiplies $R(z)$ by a factor of 2.51;
given the EDR and 2dF parameters from before, this means the new flux
ratio becomes larger than 1 for $z > 0.5$ instead of for $z > 2$.  The
tradeoff, of course, is a reduced number of total observations as well
as a smaller average optical depth to lensing for the sample.

\begin{figure}
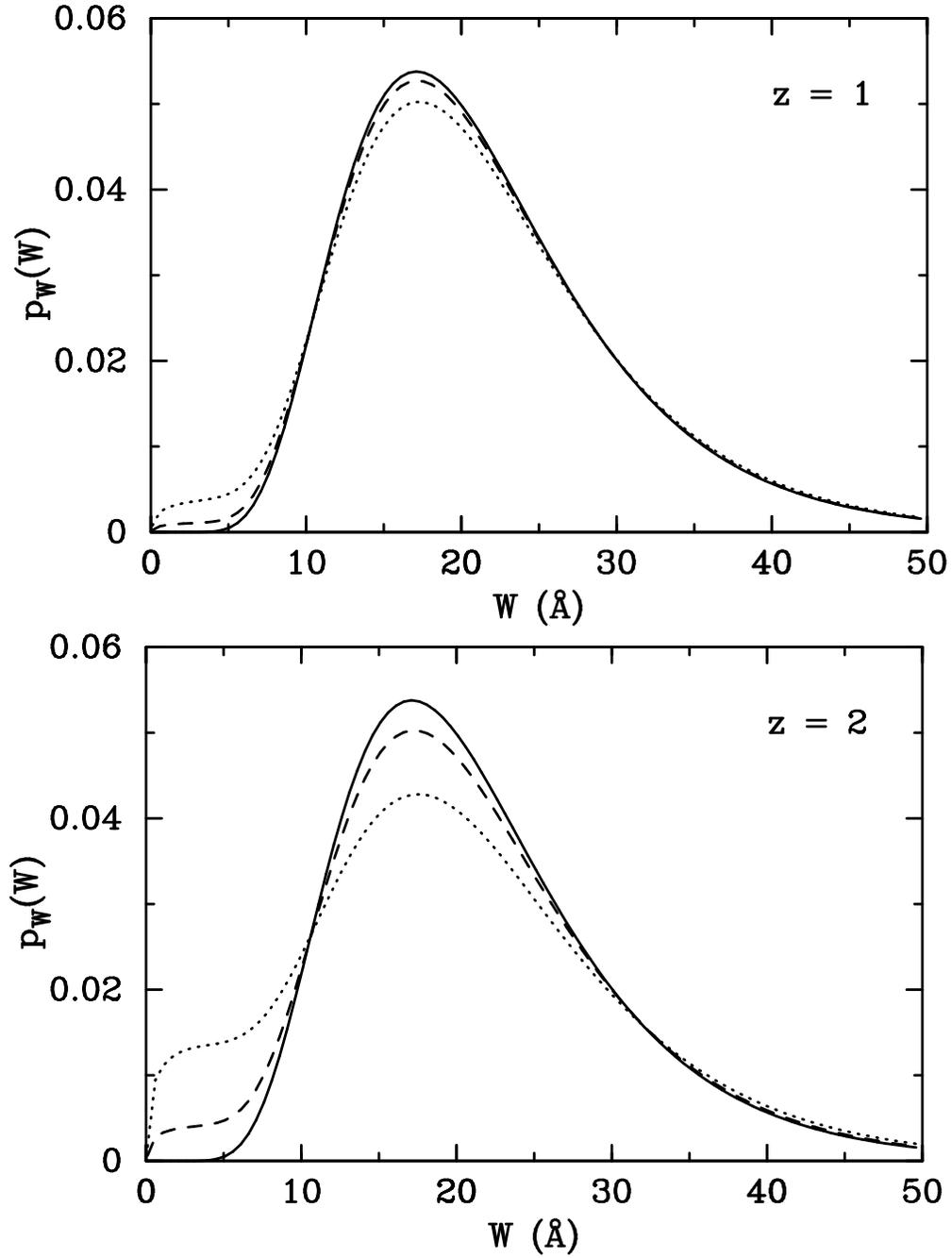

  \centering
  \epsscale{0.8}
  \plotone{f6a.eps}
  \plotone{f6b.eps}
  \epsscale{1.0}
  \caption{Lensed equivalent width distribution, given an intrinsic
  lognormal distribution, for
  $\Omega_c = 0.0$ (\emph{solid line}), 
  $\Omega_c = 0.03$ (\emph{dashed line}), and
  $\Omega_c = 0.1$ (\emph{dotted line}), 
  at redshifts of 1 (top) and 2 (bottom).}
  \label{fig:ew-dist-lognormal}
\end{figure}

While the $\delta$-function distribution is a useful beginning, a much
more realistic model for intrinsic equivalent widths is the lognormal
distribution,
\begin{equation}
  \label{eq:lognormal}
  p_0(W_0) \,dW_0 = \frac{d W_0}{\gamma \,W_0 \sqrt{2\pi}}
  e^{-(\ln W_0 - \omega)^2/2\gamma^2} .
\end{equation}
Here the ``shape parameters'' $\omega$ and $\gamma$ are the mean and
standard deviation, respectively, of $\ln W$.  This is largely an
empirically determined distribution, although there have been attempts
to provide a physical explanation \citep{qso:ew-model}.  As will be
shown in Section~\ref{cha:data}, the lognormal distribution provides a
very good approximation to the actual quasar data.
Figure~\ref{fig:ew-dist-lognormal} shows how the lensed equivalent width
model appears for various values of $z$ and $\Omega_c$, and choosing
$\omega$ and $\gamma$ as appropriate for the data set.  As in
Figure~\ref{fig:ew-dist-delta}, there is a clear lensing indicator in
the low equivalent width region that scales with both $z$ and
$\Omega_c$.

Because of the large amount of integration involved in arriving at the
final expression~(\ref{eq:pW}) for the lensed-lognormal equivalent width
distribution, it becomes necessary to compute an interpolated version
of the model.  Through a change of variables,
\begin{equation}
  \label{eq:u-of-w}
  u(W) = \frac{\ln W - \omega}{\gamma} + \gamma ,
\end{equation}
it is possible to remove $\omega$ dependence from the integral of the
lensed-lognormal model.  Then polynomial interpolation of the expression
\begin{equation}
  \label{eq:pW-xform}
  \ln [\gamma \, p_W(u(W))] + \omega - \frac{\gamma^2}{2}
\end{equation}
over the remaining four variables $(u, z, \gamma, \Omega_c)$ gives a
very good fit to the actual distribution.  In the $\Omega_c = 0$ case,
Equation~\ref{eq:pW-xform} is a simple parabola in $u$, independent of
all other parameters.

\section{Data}
\label{cha:data}

This section describes the details of the selection of the equivalent
width data set, and addresses concerns regarding the choice of an
appropriate sample to be modeled.

\subsection{SDSS Early Data Release and Quasar Catalog}
\label{sec:edr}

The Sloan Digital Sky Survey uses a dedicated 2.5~m telescope at Apache
Point Observatory, New Mexico, with an approximate $3^\circ$ field of
view, to obtain photometric and spectroscopic data at high Galactic
latitudes.  A CCD mosaic camera \citep{sdss:photo-camera} obtains
imaging data in drift-scan mode for five broad optical photometric bands
($ugriz$) that were designed for the Survey \citep{sdss:photo-filter}.
From the imaging data, up to 640 objects per field are selected for
simultaneous observation by a pair of fiber-optic spectrographs covering
the range from 3800~\AA\ to 9200~\AA.  In addition, an auxiliary 20~inch
(0.5~m) Photometric Telescope at the site provides photometric
calibrations (see \citealt{sdss:tech-summary} for more details).

The Early Data Release (EDR) \citep{sdss:edr} consists of 462~square
degrees of imaging data from eight drift scans, and 54,008 individual
spectra.  The data were acquired along the celestial Equator in both the
northern and southern Galactic skies, with additional fields
corresponding to a portion of the \emph{SIRTF} First Look Survey region.
This first public data release represents only a fraction of the planned
survey coverage of the northern ($10^4$~deg$^2$) and southern ($\sim
750$~deg$^2$) Galactic caps.  Because the SDSS photometric system was
not finalized at the time of the public data release, the EDR photometry
is referred to here as ($\uband\gband\rband\iband\zband$).

SDSS quasar candidates are selected from photometric data using a
sophisticated multicolor algorithm (see \citealt{sdss:qso-target} for
details).  Candidates for follow-up spectroscopy must also meet PSF
$\iband$ magnitude requirements.  There is a bright cutoff of $\iband >
15.0$ to prevent brighter objects' spectra from bleeding into adjacent
spectra on the CCD detector.  A Galactic extinction-corrected faint
cutoff of $\iband < 19.1$ (for $z \lesssim 3$ quasars) flux-limits the
multicolor sample; this limit was chosen to satisfy catalog
completeness requirements given the sky density of quasars and the
number of fibers allocated to quasars per spectroscopic
plate.\footnote{The 95\% completeness limit for PSF magnitudes is much
  better than this, $\iband < 21.3$ for the EDR.}  Of course,
high-redshift quasars and those targeted by other means (e.g.,
serendipity) may be fainter than this limit.  One important caveat is
that the EDR employed earlier versions of the quasar targeting algorithm
for its commissioning data, slightly modifying the magnitude limits.  In
particular, two of the image processing runs have a bright cutoff at
$\iband = 16.5$ (the remainder use $\iband = 15.0$), and for all EDR
data the faint dereddened limit is $\iband < 19.0$.

\begin{figure}
  \centering
  \plotone{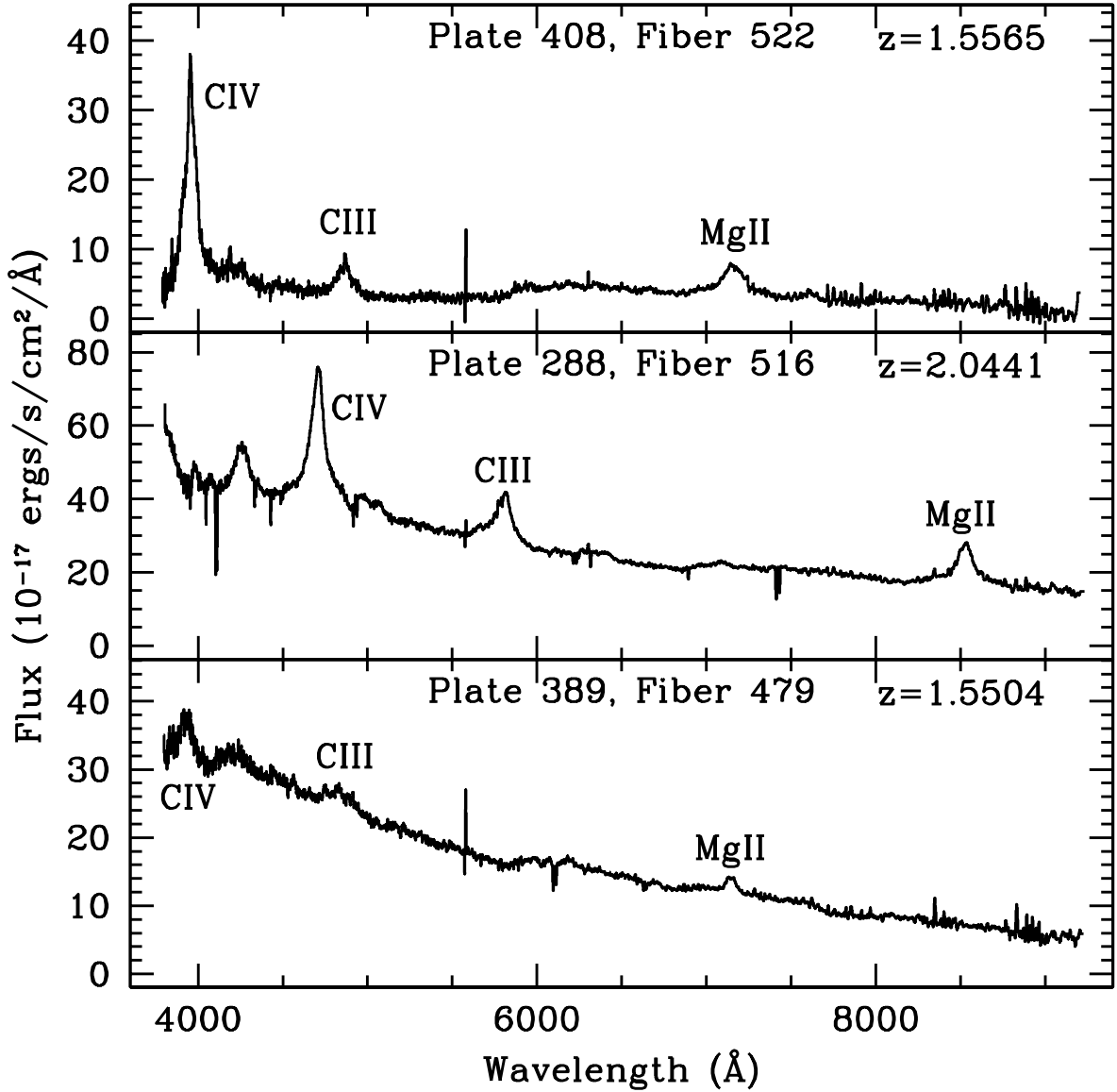}
  \caption{Sample quasar spectra at moderate redshift from the EDR,
    illustrating large equivalent widths in all three lines (top), a
    bright quasar (middle), and very low equivalent widths in all three
    lines (bottom).  The last quasar is a good microlens candidate.}
  \label{fig:qso-specs}
\end{figure}

The first edition of the SDSS Quasar Catalog \citep{sdss:edr-qsocat},
based on the EDR, includes 3814 objects, 3000 of which were discovered
by the SDSS.  Quasars with reliable redshifts were selected for
inclusion in the catalog if their spectra contained at least one broad
emission line (FWHM~$> 1000$~km~s$^{-1}$) and their luminosity exceeded
$M_{\iband} = -23$ (calculated in a $H_0 = 50$~km~s$^{-1}$~Mpc$^{-1}$,
$\Omega_M = 1.0$, $\Omega_\Lambda = 0.0$ cosmology).  The quasars range
in redshift from 0.15 to 5.03.  It is expected that the final SDSS
Quasar Catalog will contain on the order of $10^5$ quasars.  A few
sample spectra from the catalog are presented in
Figure~\ref{fig:qso-specs}.

\subsection{Creation of Equivalent Width Samples}
\label{sec:ew-samples}

For the purposes of this paper, I examined the strong \mgii,
\ciii, and \civ\ broad emission lines from EDR quasars.  The first step
was to extract information from the SDSS Catalog Archive using the SDSS
Query Tool.  I selected all measured spectral lines belonging to objects
classified as quasars, and for which the redshift measurements were not
deemed to be of low confidence.  A sample query for the \mgii\ line, in
SQL notation, is
\begin{verbatim}
  SELECT spec.plate.plateID, spec.plate.mjd, spec.fiberID
  FROM sxSpecLine
  WHERE (category == 1 && name.lineID == 2800 &&
    (spec.specClass == SPEC_QSO || spec.specClass == SPEC_HIZ_QSO) &&
    (spec.zWarning & Z_WARNING_LOC) == 0)
\end{verbatim}
This first set of queries resulted in 2809 \mgii\ lines, 2295 \ciii\ 
lines, and 1650 \civ\ lines, from a total of 3280 spectroscopic
observations classified as quasars in the EDR.  (Without the
low-confidence redshift constraint, these numbers are 3368, 2738, and
1909 respectively, from 3925 quasars.)

Next, I removed from the results of the Archive queries all objects that
did not appear in the official SDSS Quasar Catalog.  This reduced the
number of distinct quasars to 3095 (2637 \mgii\ lines, 2215 \ciii\ 
lines, and 1595 \civ\ lines).  There are 130 objects in the Quasar
Catalog not tagged as low-confidence that do \emph{not} appear in the
any of catalogs returned by the SQL queries.  Of these, the vast
majority have redshifts either too small or too large for the spectral
lines of interest to appear in the spectroscopic data.  The remainder
include 7 of the 10 objects added to the catalog from a visual
inspection of EDR spectra, and 14 of the 16 objects added during a
visual search for extreme broad absorption-line (BAL) quasars.

\begin{figure}
  \centering
  \epsscale{0.9}
  \plotone{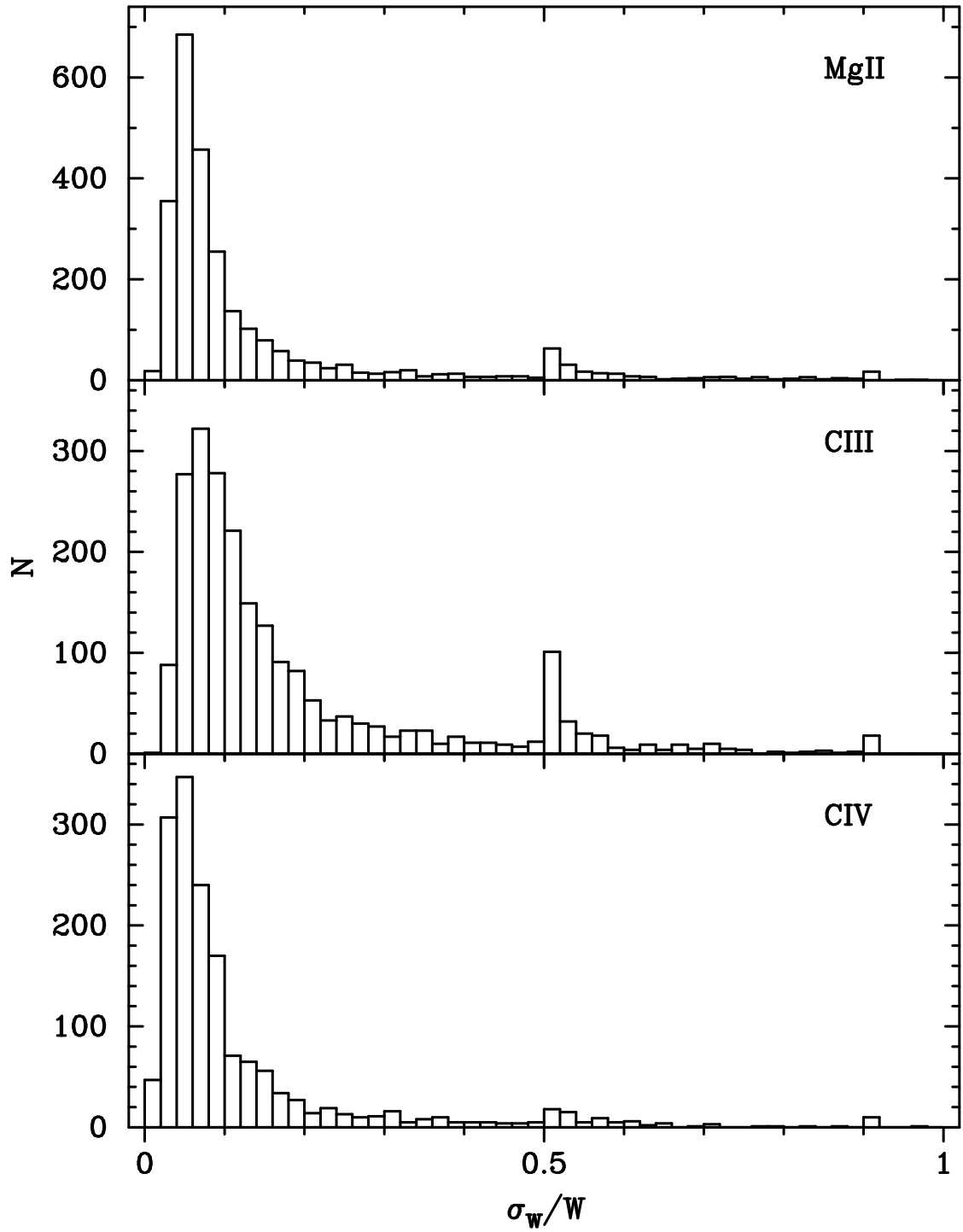}
  \epsscale{1.0}
  \caption{Histograms of equivalent width fractional error for \mgii,
  \ciii, and \civ\ respectively.}
  \label{fig:ewfrac-hist}
\end{figure}

\begin{figure}
  \centering
  \epsscale{0.9}
  \plotone{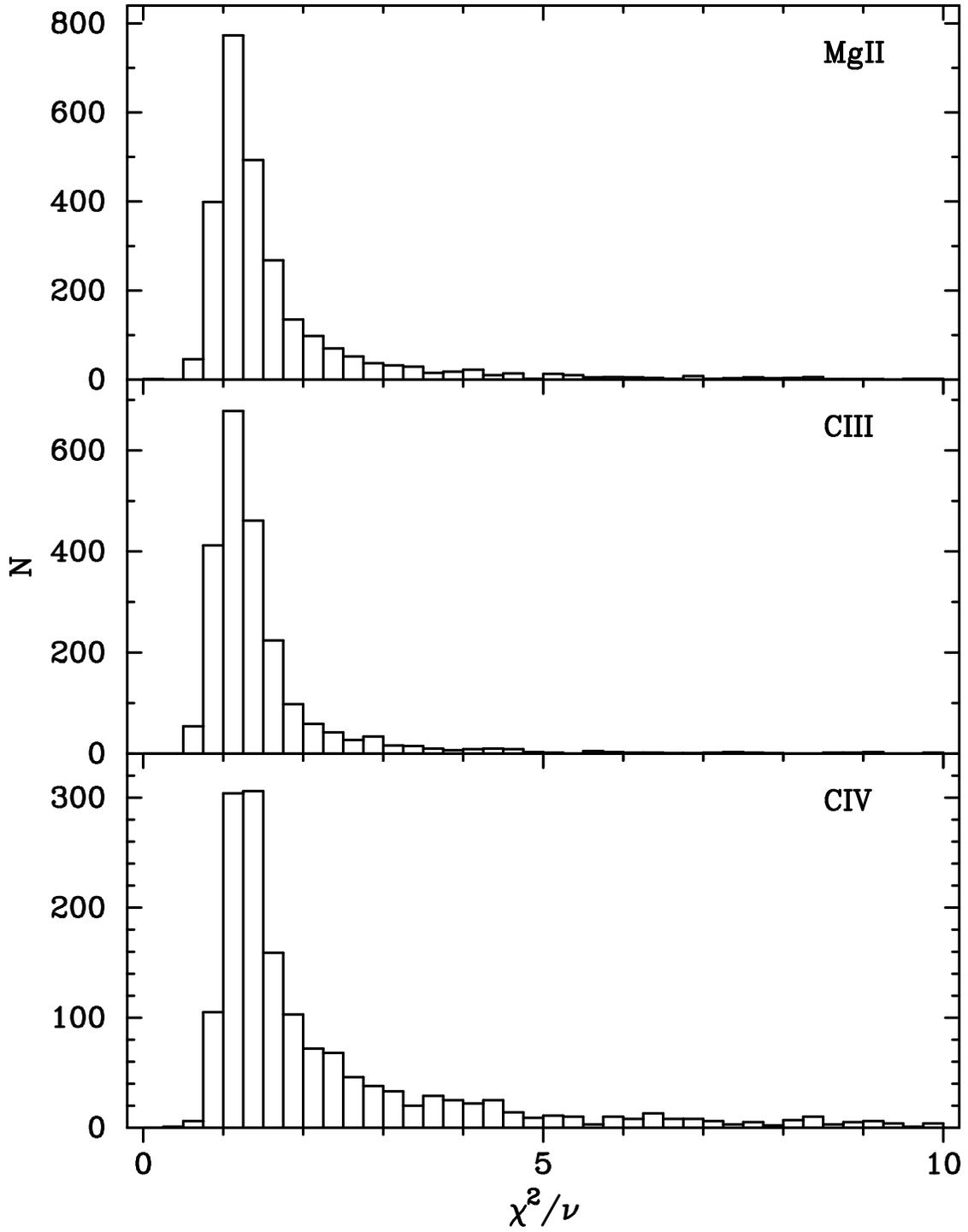}
  \epsscale{1.0}
  \caption{Histograms of $\chi^2/\nu$ for \mgii, \ciii, and \civ\
  respectively.}
  \label{fig:chisq-hist}
\end{figure}

Spectral line parameters are determined by fitting a single Gaussian to
the con\-tinuum-sub\-tracted flux in the neighborhood of an identified
line.  Several entries in the Catalog Archive describe the quality of
the spectral line fits.  In particular, I considered the fractional
error of the equivalent width, $\sigma_W/W$ (calculated from the
uncertainties in the fit parameters), and the chi-squared per degree of
freedom for the fit, $\chi^2/\nu$.  Histograms of these parameters for
each data set are shown in Figures~\ref{fig:ewfrac-hist}
and~\ref{fig:chisq-hist}.  There are anomalous spikes in the equivalent
width fractional error histograms just above 0.5 and again at 0.9, for
reasons that remain unclear.

\begin{figure}
  \centering
  \epsscale{0.8}
  \plotone{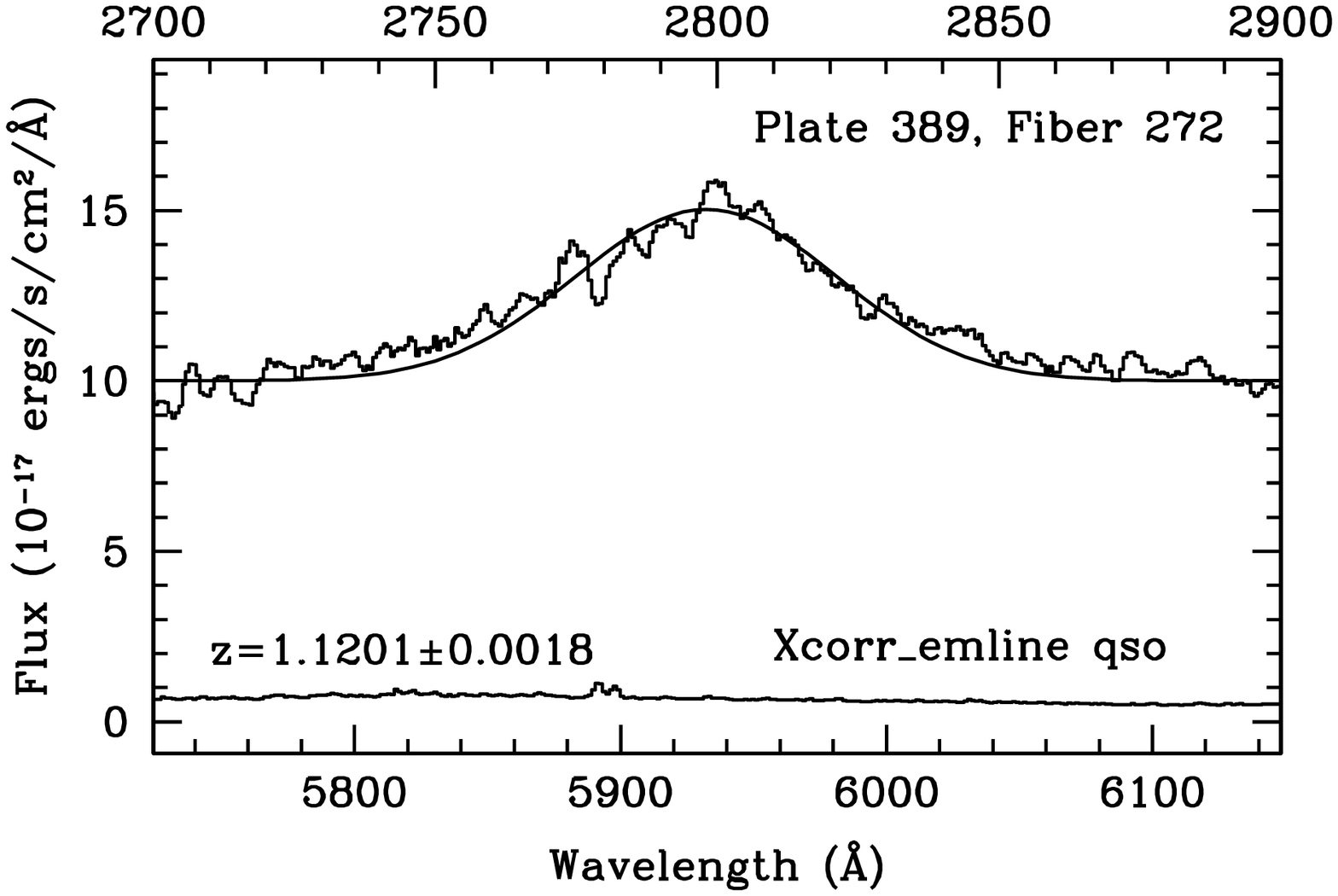}
  \plotone{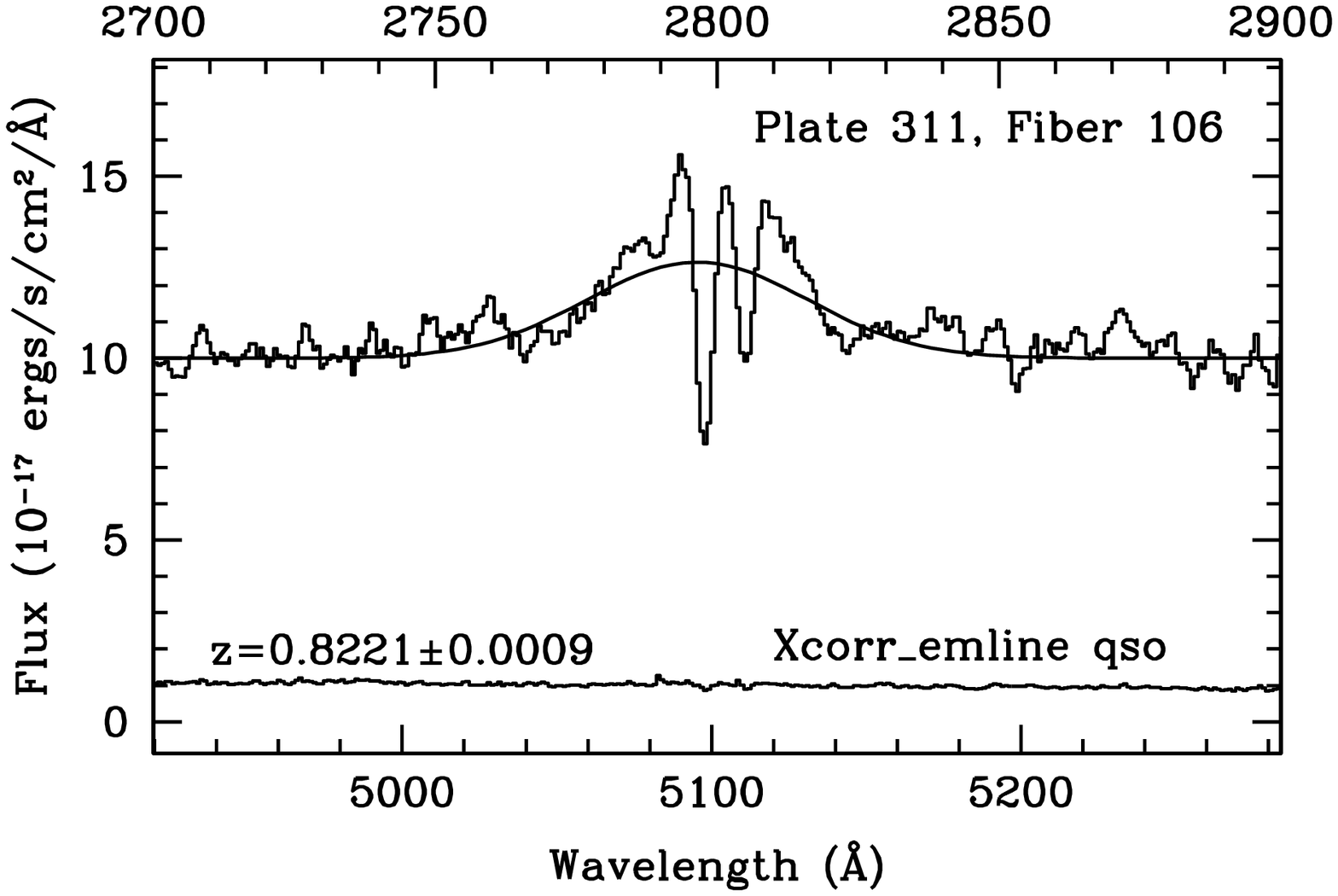}
  \epsscale{1.0}
  \caption{Examples of good (top) and poor (bottom) Gaussian line fits
    to \mgii.  Each plot shows the best fit to the continuum-subtracted
    flux, with the error displayed beneath.  The line in the bottom plot
    is marred by an absorption system.}
  \label{fig:sample-fits}
\end{figure}

Based on these histograms, I chose a reasonable absolute minimum ``data
quality'' standard of $\sigma_W/W \leq 0.5$, $\chi^2/\nu \leq 5.0$, with
an additional requirement that the equivalent width itself be a positive
number.  In Section~\ref{cha:analysis} I consider the effect that more
restrictive data quality cuts in $\sigma_W/W$ and $\chi^2/\nu$ have on
the constraints placed on $\Omega_c$.  This reduced the data set to 2333
\mgii\ lines, 1929 \ciii\ lines, and 1322 \civ\ lines.  I then visually
inspected all spectral lines and their fits, eliminating observations
with poor fits due to large nearby absorption systems, excessive noise
from atmospheric lines, and other examples of poor fitting on the part
of the spectroscopic pipeline algorithms; see
Figure~\ref{fig:sample-fits} for sample line fits.  In general I opted
to keep truly borderline observations, reasoning that these data would
tend to make the bounds on $\Omega_c$ more conservative, and also with
the knowledge that many such borderline objects would be eliminated
automatically by stricter data quality cuts.  In all, 99 \mgii\ lines,
61 \ciii\ lines, and 61 \civ\ lines were removed from the data sample
based on visual inspection.

\begin{figure}
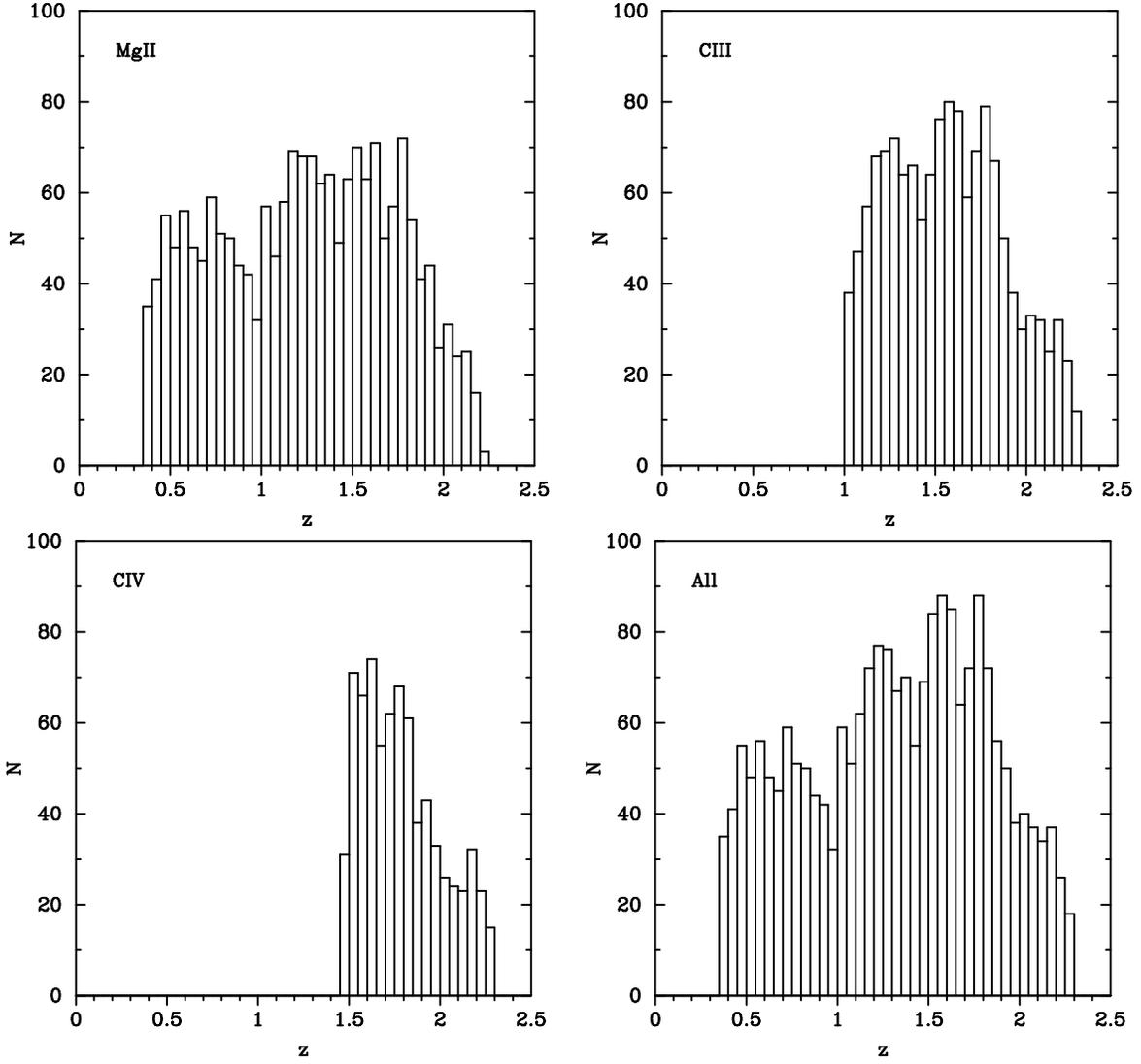

  \centering
  \epsscale{0.45}
  \plotone{f11a.eps}\hfil\plotone{f11b.eps}\\
  \plotone{f11c.eps}\hfil\plotone{f11d.eps}
  \epsscale{1.0}
  \caption{Histogram of redshift for each spectral line, as well as for
    all quasars in the sample.}
  \label{fig:zhist}
\end{figure}

\begin{figure}
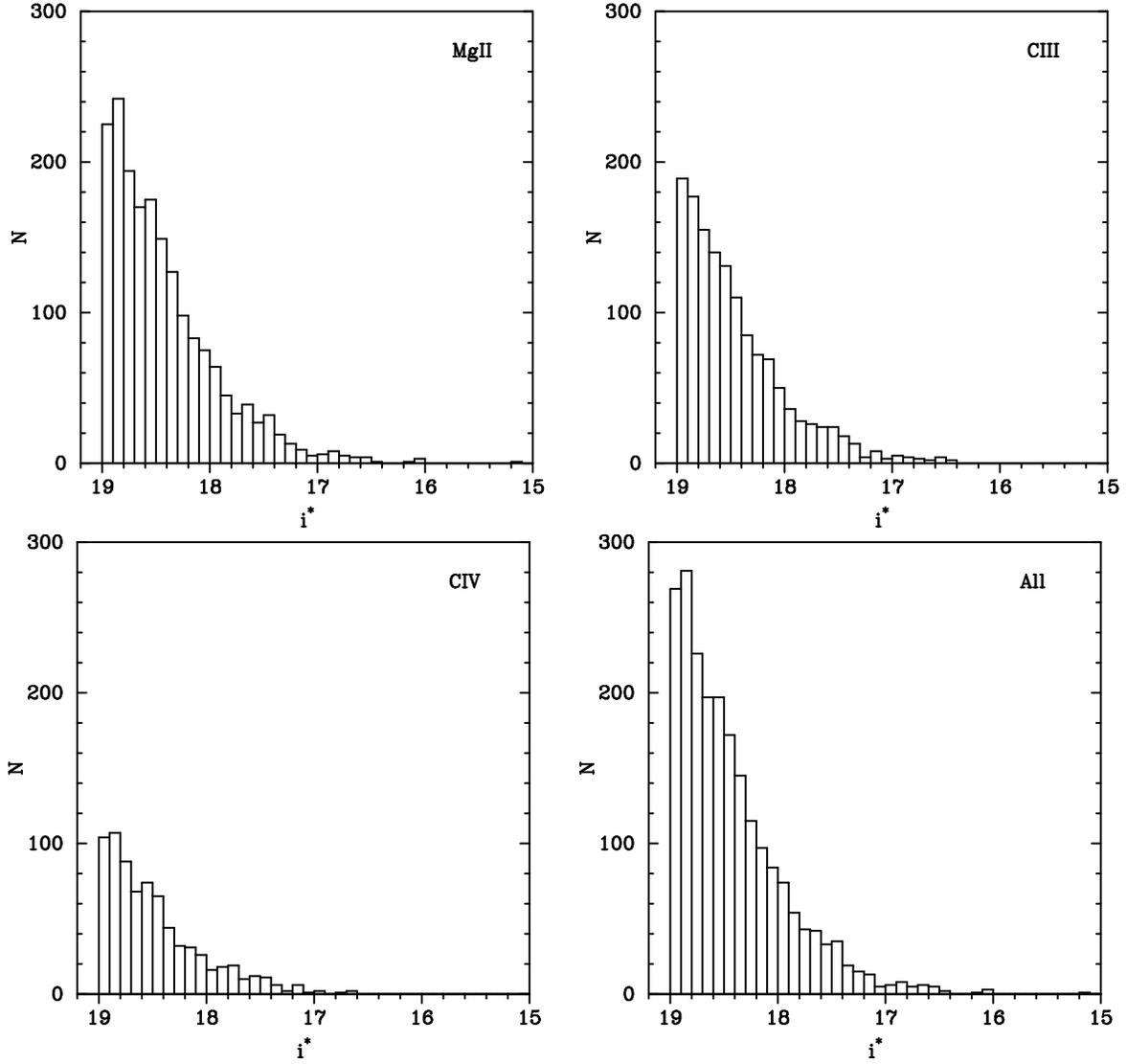

  \centering
  \epsscale{0.45}
  \plotone{f12a.eps}\hfil\plotone{f12b.eps}\\
  \plotone{f12c.eps}\hfil\plotone{f12d.eps}
  \epsscale{1.0}
  \caption{Histogram of dereddened $\iband$ magnitude for each spectral
  line, as well as for all quasars in the sample.}
  \label{fig:imaghist}
\end{figure}

Finally, based on the redshift limits of the 2dF luminosity function ($z
\leq 2.3$), and the quasar targeting algorithm faint limit (dereddened
$\iband < 19.0$), I obtain 1857 \mgii\ equivalent widths with $0.36 < z
< 2.21$, 1382 \ciii\ observations with $1.01 < z < 2.3$, and 745 \civ\ 
equivalent widths with $1.47 < z < 2.3$.  Figures~\ref{fig:zhist}
and~\ref{fig:imaghist} show the number of observations of each line
binned by redshift and dereddened $\iband$ apparent magnitude.

\subsection{Appropriateness of Sample}
\label{sec:sample-apropos}

There are several important considerations that apply to the choice of
quasar sample for this test.  One central assumption in the development
of the equivalent width model distribution is that the sample is
continuum flux-limited, in order that amplification bias may be treated
correctly.  Optically selected samples such as the SDSS Quasar Catalog
will not be exactly limited by continuum flux due to contributions from
line emission.  Because the sample is limited by $\iband$ magnitude, the
\mgii\ line will contribute to the measured filter flux for quasars with
redshifts $1.4 < z < 1.9$ \citep{sdss:photo-filter}; the \ciii\ and
\civ\ lines not appear in $i$ over the redshifts studied, although other
lines such as \oiii\ and some of the Balmer series do move into the $i$
filter at low redshift.  However, the width of the filter is about
1200\AA\ while the typical observed emission line equivalent width is
less than 10\% of this, so the typical magnitude in this situation will
be less than $0.1$ brighter than the continuum-only value.

It is also important for the sample to be as complete as possible.  The
SDSS quasar targeting algorithm \citep{sdss:qso-target} has been shown
to be at least 90\% complete for $z < 2.5$ and $\iband < 19.1$ (19.0 for
the EDR).  Since the luminosity function is only determined for $z <
2.3$ \citep{twodf:lf}, completeness in redshift is not a concern.  The
Quasar Catalog absolute magnitude limit of $M_\iband = -23$ does not in
practice affect the sample either.  As shown in Figure~3 of
\citet{sdss:edr-qsocat}, this restriction removes only \mgii\ 
observations near $\iband = 19.0$ with $z < 0.42$, a negligibly small
population that is highly unlikely to exhibit microlensing.  The Quasar
Catalog requirement that FWHM~$> 1000$~km~s$^{-1}$ for at least one
emission line is more worrisome, since this may exclude small equivalent
width observations.  On the other hand, this criterion was chosen
specifically to eliminate narrow-lined quasars, which don't belong in a
study of the broad emission line region.  Moreover, only 10 quasars were
dropped from the catalog based solely on this criterion.  Finally, the
bright-limit discrepancy ($\iband = 15.0$ versus $\iband = 16.5$) in
different EDR processing runs affects only a very few observations, as
can be seen from the relative paucity of bright quasars in
Figure~\ref{fig:imaghist}.

\subsubsection{Emission Lines}
\label{sec:em-lines}

The microlensing signature in the equivalent width distribution depends
on the fact that a broad emission line arises from a much larger region
than does the continuum surrounding it in the spectrum.  Although \mgii\ 
can be considered the most useful emission line due to both the wide
range of redshift and sheer numbers, it suffers from being located in
the midst of what has been termed the ``small blue bump.''  What appears
to be a slightly brighter region of the rest-frame continuum from
2000-3600\AA\ is in fact the continuum plus blended \feii\ multiplets
and high-order Balmer series lines \citep{qso:blr-feii}.  If the \mgii\ 
equivalent width is calculated relative to this false high continuum (as
in the SDSS spectroscopic pipeline), any magnification of the true
continuum due to lensing will be underestimated.  The composite SDSS
quasar spectrum of \citet{sdss:qso-composite} suggests that the
contribution of line emission to the apparent continuum around \mgii\ is
on the order of 25-50\%.  Thus high-magnification lensing events might
appear to have an effective magnification only $2/3$ the actual value.
This would tend to dilute the lensing signal, leading to overly
optimistic constraints on $\Omega_c$.

The \ciii\ and \civ\ lines do not share the continuum problems of \mgii.
Because they enter the optical region at higher redshifts, however, they
lack the advantage of a long redshift baseline.  There are also
necessarily fewer observations in the flux-limited sample.  \ciii\ does
have one edge over the others; line variability studies suggest that
\ciii\ may be emitted from larger regions of the quasar than other lines
\citep{book:agn-peterson}.

\subsubsection{Time Dependence of Microlensing}
\label{sec:time-dep}

While the characteristic time dependence of microlensing is the key to
many of its applications, such is not the case with this method.  It is
important that the spectroscopic observations not lag far behind the
photometric imaging.  If this delay is too large, the microlensing event
that may have amplified a quasar into the flux-limited sample will have
ended, and the equivalent widths will no longer reflect the fact that
the quasar was once lensed.  To estimate the decorrelation timescale,
consider the time it takes for a lens moving with velocity $v_\perp$
(perpendicular to the line of sight) to cover a distance equal to its
Einstein radius \citep{book:sef}:
\begin{equation}
  \label{eq:timescale}
  \Delta t = \frac{\alpha_E D_L}{v_\perp} \approx 78 h^{-1/2} 
  \left( \frac{300 \mbox{ km s$^{-1}$}}{v_\perp}\right)
  \left( \frac{M}{\Msolar} \right)^{1/2} \mbox{yr} ,
\end{equation}
where $v_{rel}$ is the velocity of the lens perpendicular to the line of
sight.  For a conservative estimate of a $0.001 \Msolar$ lens with a
velocity dispersion of $300$~km~s$^{-1}$, this timescale is only
2.9~yr.

\begin{figure}
  \plotone{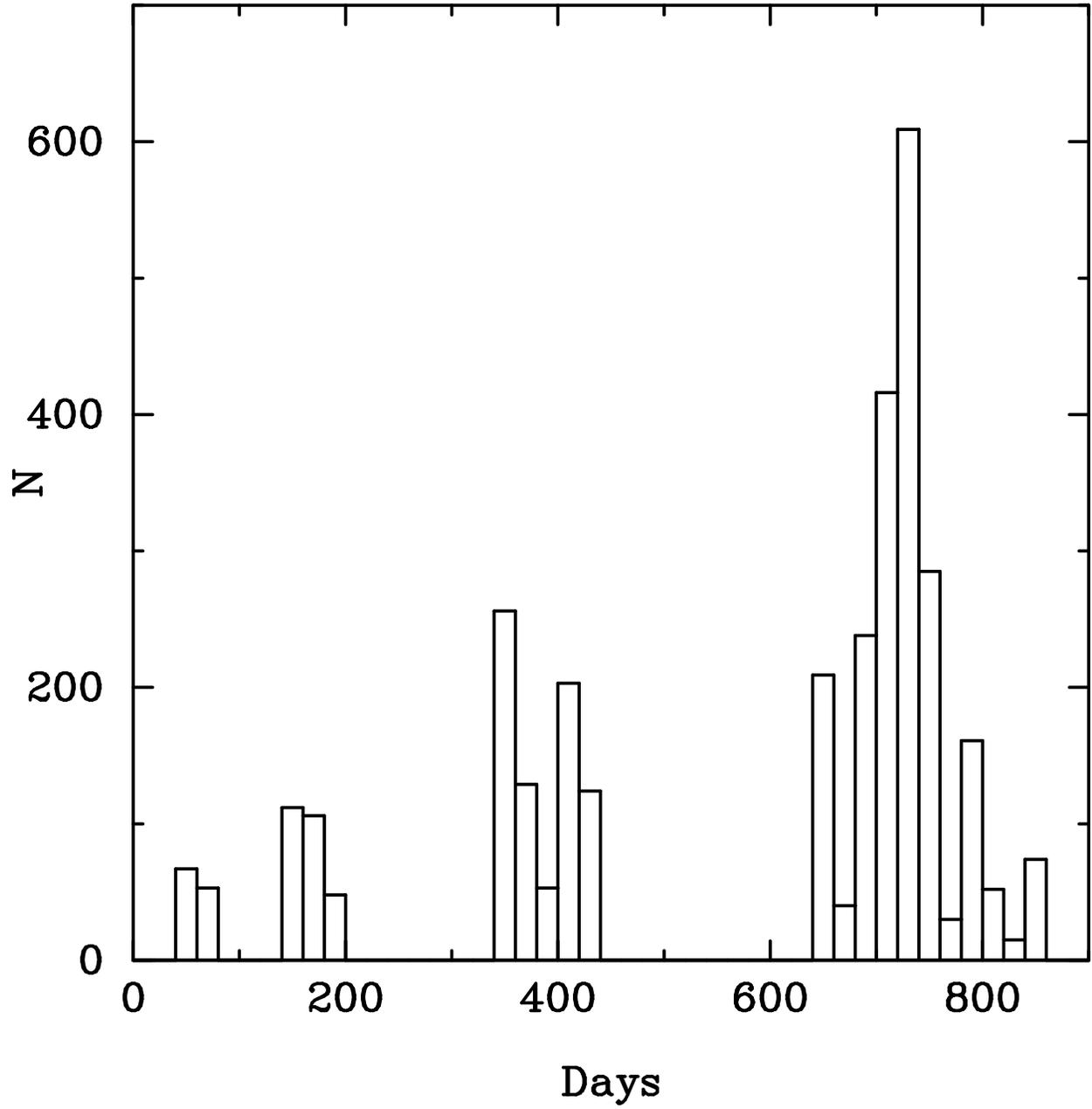}
  \caption{Difference in MJD between photometric and spectroscopic
    observations for EDR quasars.}
  \label{fig:mjd-diffs}
\end{figure}

To address this concern, I queried the SDSS Catalog Archive for the MJD
of the spectroscopic and photometric observations of the quasars.  The
median time delay between observations was 1.9~yr, with a minimum of
just 0.14~yr and a maximum of 2.3~yr; Figure~\ref{fig:mjd-diffs} shows a
histogram of the time delays.  While these time differences are smaller
than the timescale estimate above, they are a potential source of worry
for the EDR data.  It is however important to note that an order of
magnitude larger lens mass pads the decorrelation timescale by a
comfortable factor of 3.  Moreover, because the quasar survey flux limit
is for the most part less than the characteristic flux of the
luminosity function (see Figure~\ref{fig:Rz}), the amplification bias is
correspondingly weaker than if the flux limit resided on the steep
power-law slope of the LF.  The fewer observations there are due to
bias, the less the decorrelation between photometric and spectroscopic
measurements matters.

\subsubsection{Other Quasar Samples}
\label{sec:qso-others}

The first EDR-based SDSS Quasar Catalog is by no means the only
flux-limited sample available, nor is it the largest.  The Large Bright
Quasar Survey \citep*{lbqs:qsocat} set the standard for the current era
of massive spectroscopic surveys, containing approximately 1/4 the
number of quasars in the EDR.  The 2dF collaboration has far surpassed
both with a publicly released spectroscopic catalog of $10^4$ quasars
\citep{twodf:10kcat}, and will soon release the full catalog of nearly
24,000 quasars.\footnote{See \url{http://www.2dfquasar.org/}.}  The
SDSS also very recently announced its ``beta version'' of Data Release
1, containing 17,700 quasar spectra with $z < 2.3$.\footnote{See
  \url{http://www.sdss.org/dr1/} for the pipeline data products.  A
  relational database catalog will also be made available in the near
  future.}

Despite the smaller size, the EDR Quasar Catalog edges out the 2dF
catalog in several ways.  The SDSS spectrograph obtains data to longer
wavelengths (9200~\AA\ compared to 2dF's 7900~\AA\ limit), allowing higher
redshift observations of the \mgii\ line in particular.  The SDSS
catalog is also 90\% complete to a slightly higher redshift than 2dF.
Perhaps most importantly, the SDSS spectroscopic data-processing
pipeline produces as a matter of course a wealth of information
characterizing the spectra and their emission lines.

\section{Analysis}
\label{cha:analysis}

This section describes the use of the maximum-likelihood method for
determining the bounds on $\Omega_c$.  Given a probability distribution
of equivalent widths (Equation~\ref{eq:pW}), the logarithm of the
likelihood function may be written generically as
\begin{equation}
  \label{eq:logl}
  \ln L(W_i; z_i, \theta) = \sum_i \ln p_W(W_i; z_i, \theta) ,
\end{equation}
where $\theta$ represents the parameter collection.  So far in the
discussion of the lensed equivalent width distribution, the parameters
consist of $(\Omega_c, \omega, \gamma )$ (i.e., the compact dark matter
fraction plus the two shape parameters of the intrinsic lognormal
distribution).  Also implicit in this list from consideration of the
extended source is $M_{lens}$; this parameter will be treated separately
below.

\begin{figure}
  \centering
  \epsscale{0.9}
  \plotone{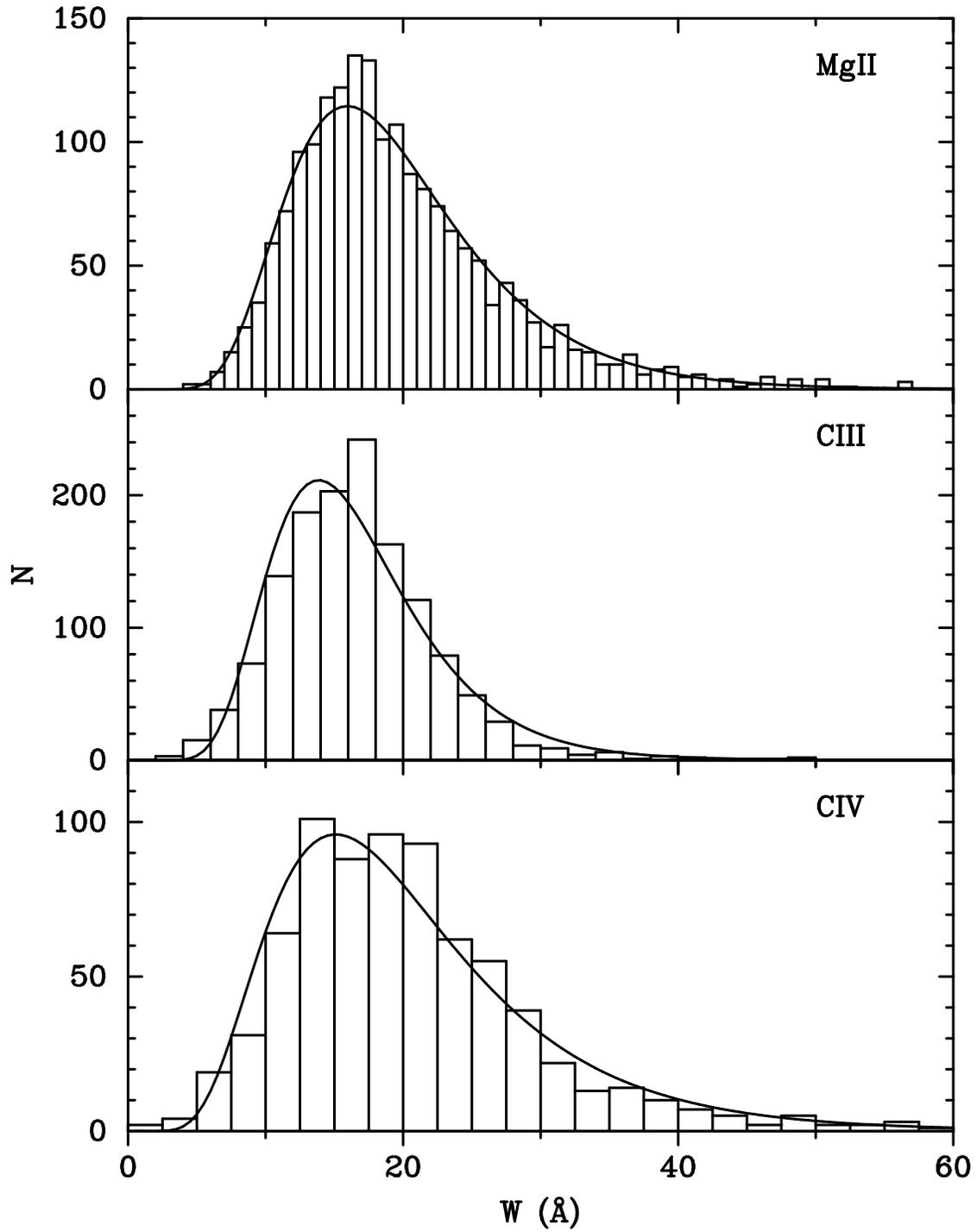}
  \epsscale{1.0}
  \caption{Equivalent width histograms for \mgii\ (top), \ciii\
  (middle), and \civ\ (bottom) data sets, along with their best-fit
  lognormals.}
  \label{fig:ewhist-all}
\end{figure}

Figure~\ref{fig:ewhist-all} shows the equivalent width distributions for
the three sets of emission line data constructed in
Section~\ref{cha:data}.  Also plotted are the best-fit lognormal
distribution for each data set.  These plots clearly demonstrate that
the model assumption of lognormality is well-motivated.

\begin{figure}
  \centering
  \epsscale{0.9}
  \plotone{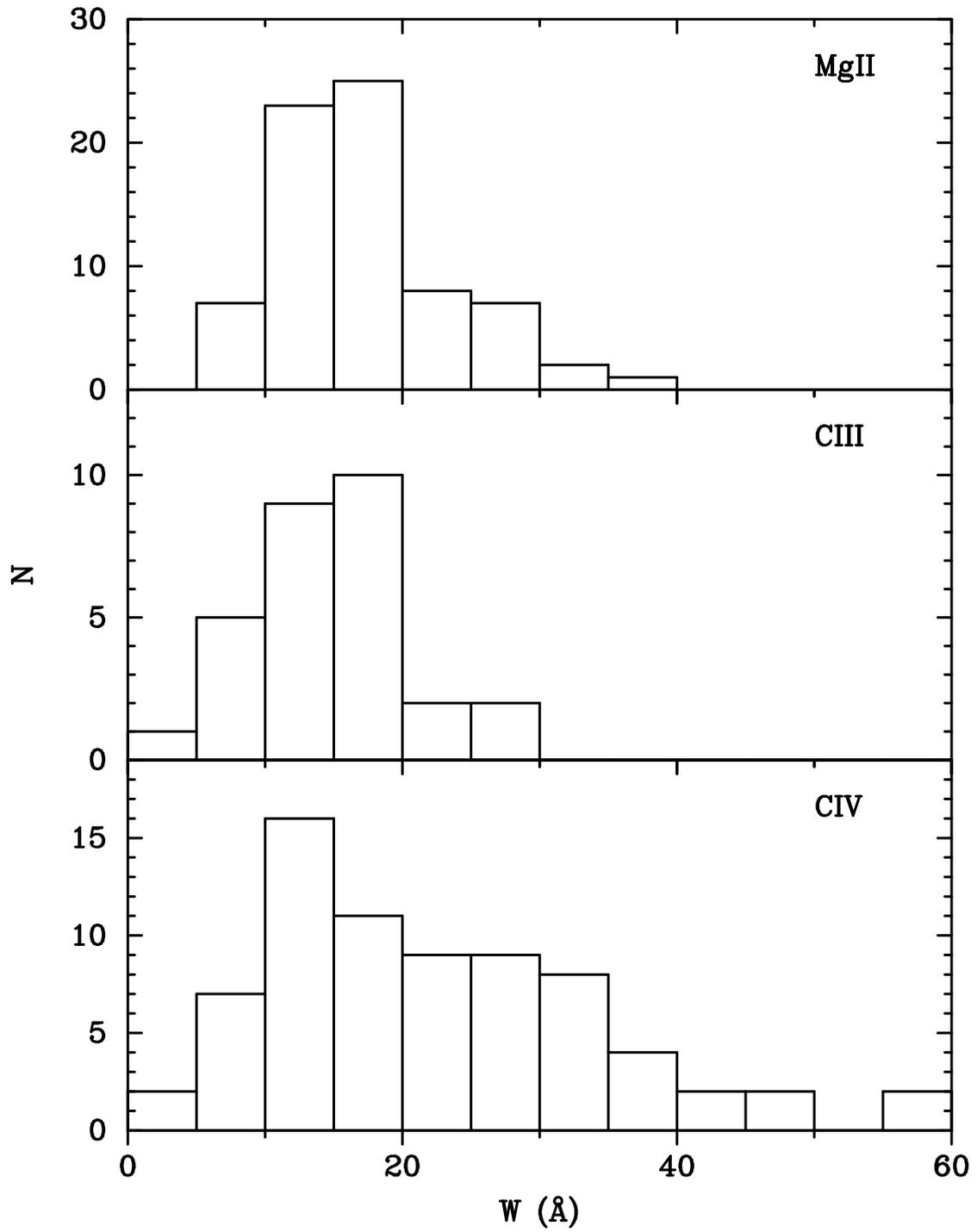}
  \epsscale{1.0}
  \caption{Equivalent width histograms for the data culled by the
    criteria $\sigma_W/W \leq 0.4$ and $\chi^2/\nu \leq 3.0$.  Once
    again, \mgii\ is on top, followed by \ciii\ and \civ.}
  \label{fig:ewhist-culled}
\end{figure}

One might worry that more restrictive ``data quality'' standards
than those imposed in Section~\ref{sec:ew-samples} (such as are
considered below) would tend to bias the data strongly for or against
evidence of lensing.  If that were the case, a histogram of the
equivalent widths culled by the new requirement ought to show evidence
of being skewed toward high or low values.
Figure~\ref{fig:ewhist-culled} shows those data removed by the criteria
$\sigma_W/W \leq 0.4$ and $\chi^2/\nu \leq 3.0$.  There is no 
indication from these histograms of obvious bias in the selection
criteria.

\begin{figure}
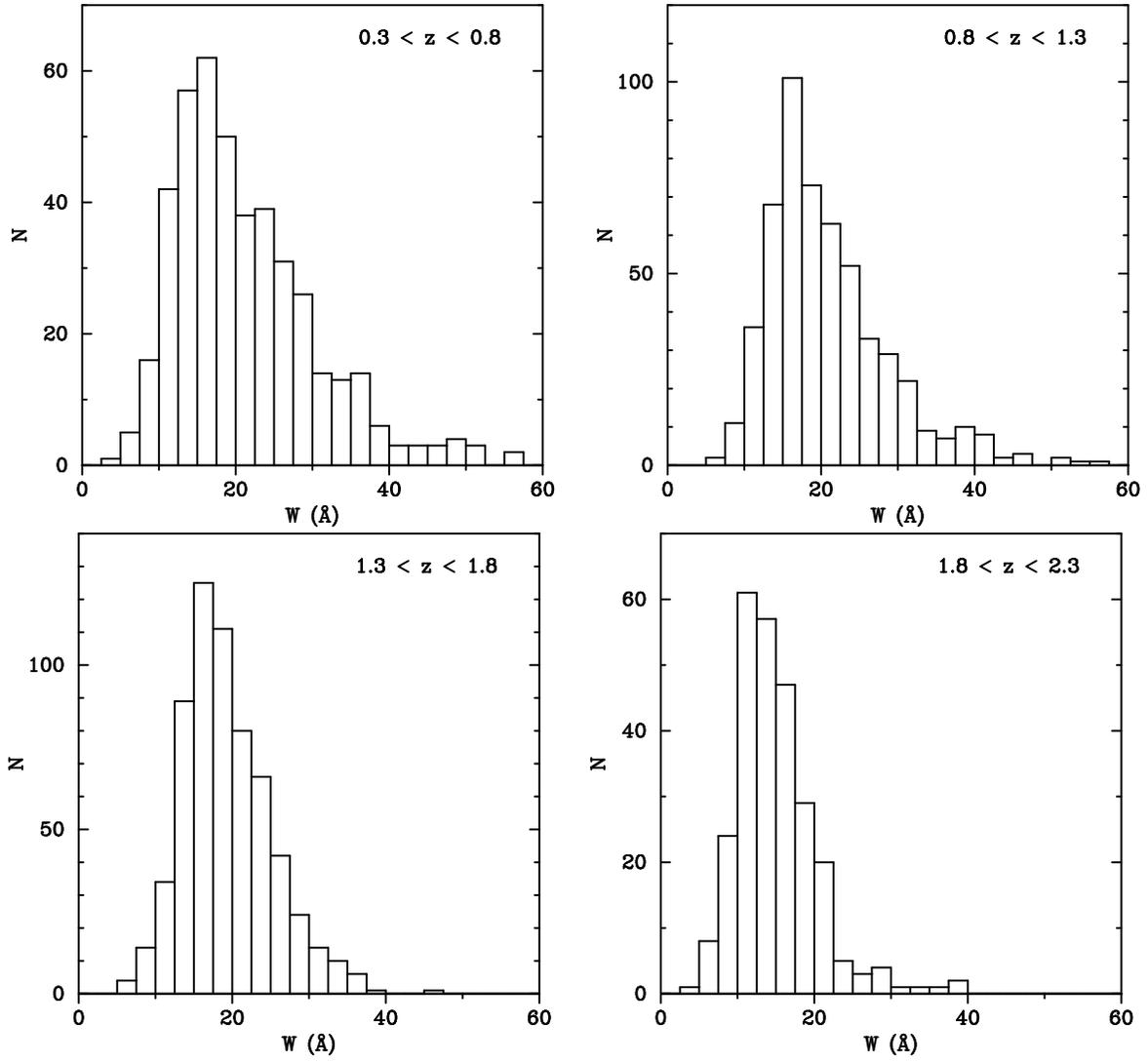

  \centering
  \epsscale{0.45}
  \plotone{f16a.eps}\hfil\plotone{f16b.eps}\\
  \plotone{f16c.eps}\hfil\plotone{f16d.eps}
  \epsscale{1.0}
  \caption{Equivalent width histograms for \mgii, for different redshift
    ranges.}
  \label{fig:ewhist-zbins}
\end{figure}

It is also instructive to divide the equivalent width data into
different redshift ranges, looking for obvious signs of microlensing.
Figure~\ref{fig:ewhist-zbins} shows this for the \mgii\ observations.
It is immediately evident by comparing these histograms to the models in
Figure~\ref{fig:ew-dist-lognormal} that the lensing effect, if any, is
quite small.  There is no apparent sign of an enhanced fraction of low
equivalent width quasars that appears to increase with redshift.

\subsection{Baldwin Effect}
\label{sec:baldwin}

Unfortunately, as Figure~\ref{fig:ewhist-zbins} illustrates, the
universe is not so kind as to oblige with a non-evolving intrinsic
equivalent width distribution requiring only two parameters.  In fact,
using just the three parameters $(\Omega_c, \omega, \gamma )$ listed
above yields a very poor fit to the EDR quasar data, as measured by a
$\chi^2$ statistic ($\chi^2/\nu \gtrsim 2.3$ for \mgii).  The difficulty
is due to a phenomenon known the Baldwin effect, in which the equivalent
width of many quasar broad emission lines is anti-correlated with
luminosity \citep{bald:baldwin,twodf:baldwin} or perhaps, as some have
suggested, with redshift \citep*{bald:lbqs}.

\begin{figure}
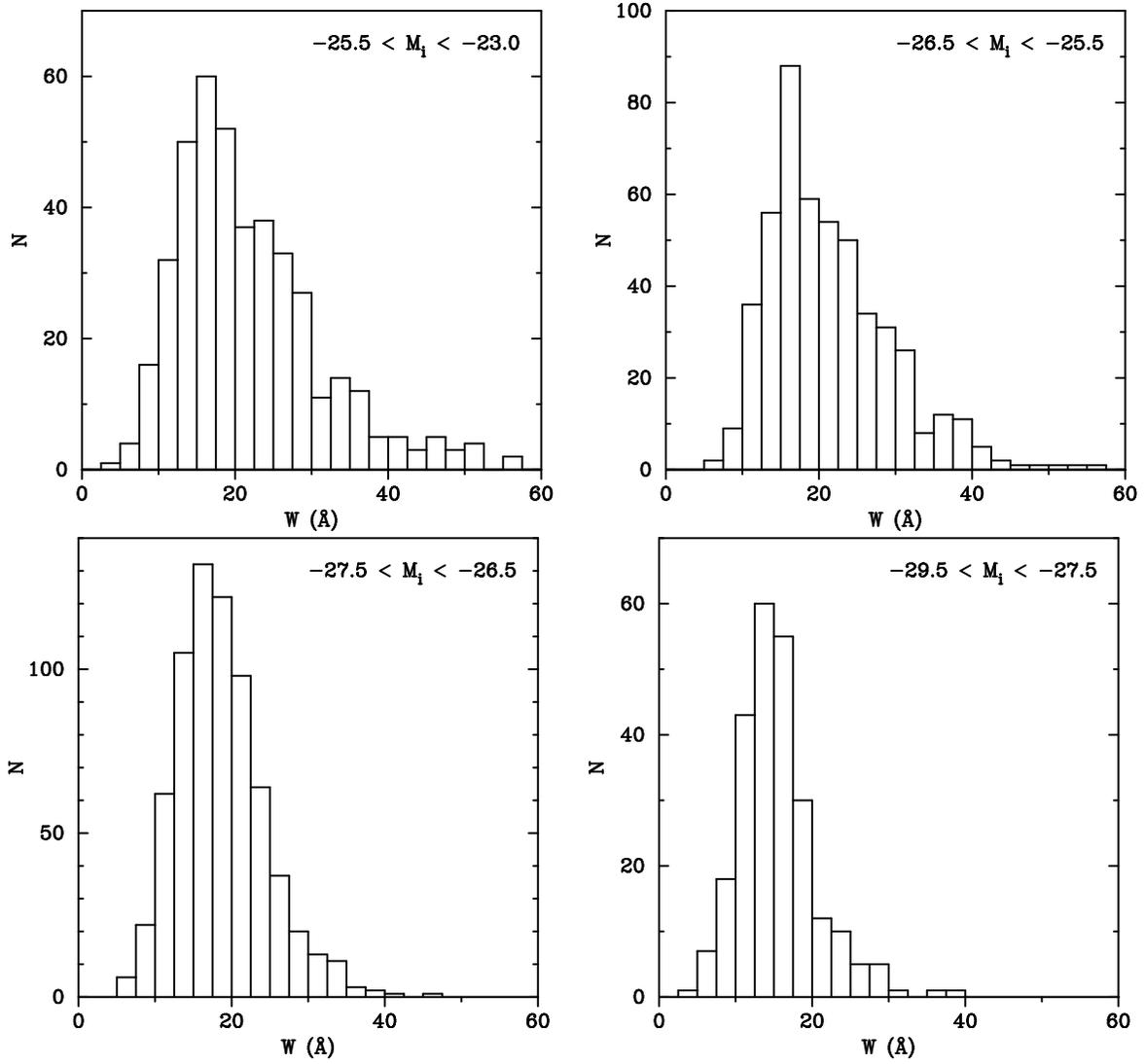

  \centering
  \epsscale{0.45}
  \plotone{f17a.eps}\hfil\plotone{f17b.eps}\\
  \plotone{f17c.eps}\hfil\plotone{f17d.eps}
  \epsscale{1.0}
  \caption{Equivalent width histograms for \mgii, for different absolute
    magnitude ranges.}
  \label{fig:ewhist-magbins}
\end{figure}

While Figure~\ref{fig:ewhist-zbins} would seem to indicate a
redshift-dependent evolution of the equivalent width distribution, this
could be a result of the fact that higher redshift quasars in a
flux-limited sample will tend to have higher luminosity.
Figure~\ref{fig:ewhist-magbins} shows histograms of the observations in
different ranges of absolute magnitude in $\iband$, and suggests a
similar trend.

\begin{figure}
  \centering
  \epsscale{0.45}
  \plotone{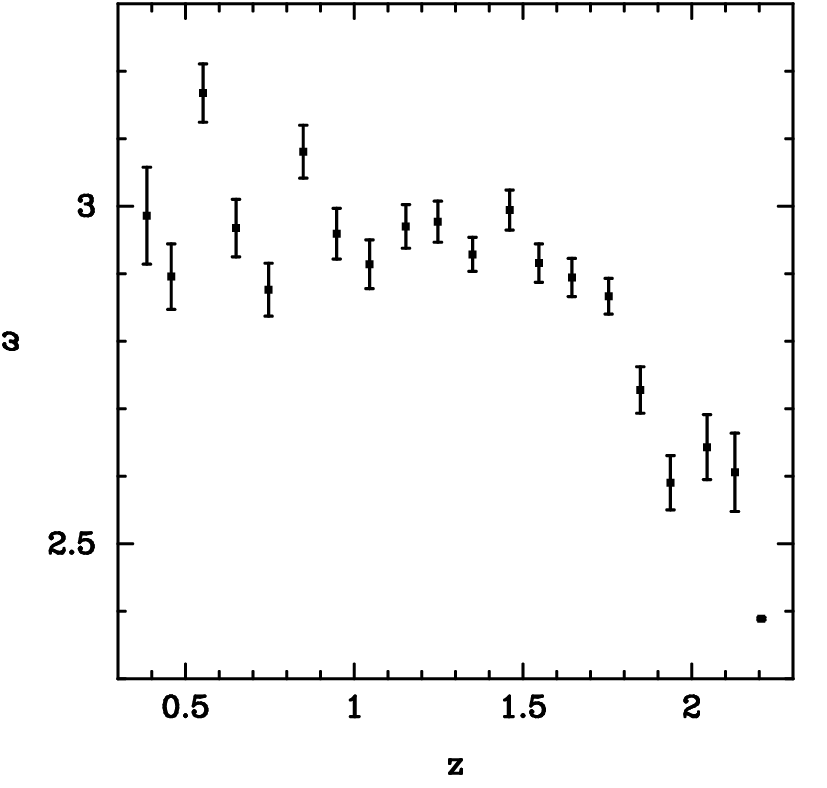}\hfil\plotone{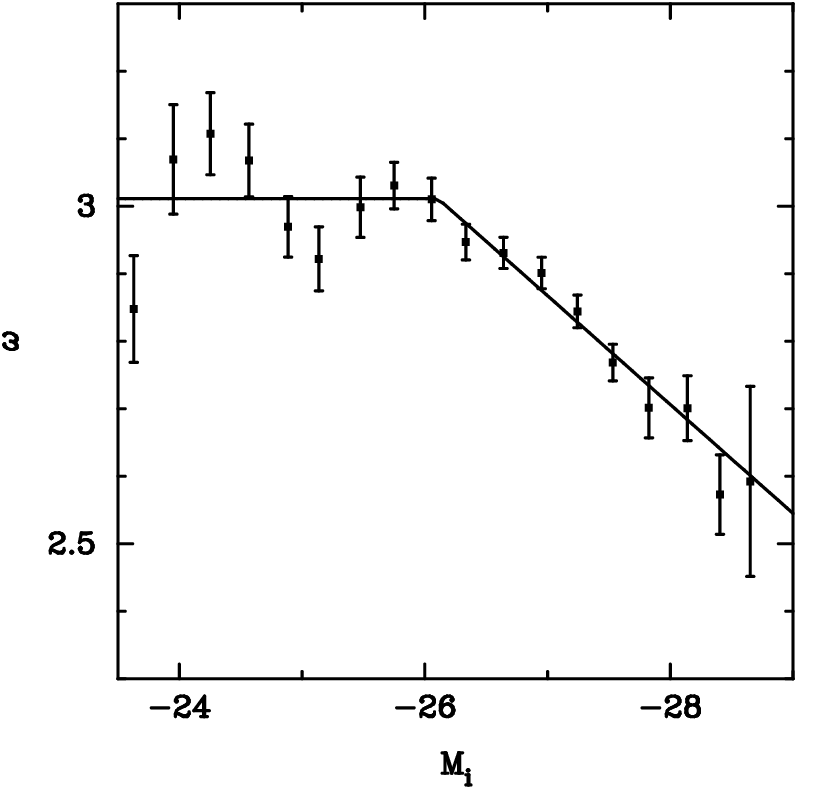}\\
  \plotone{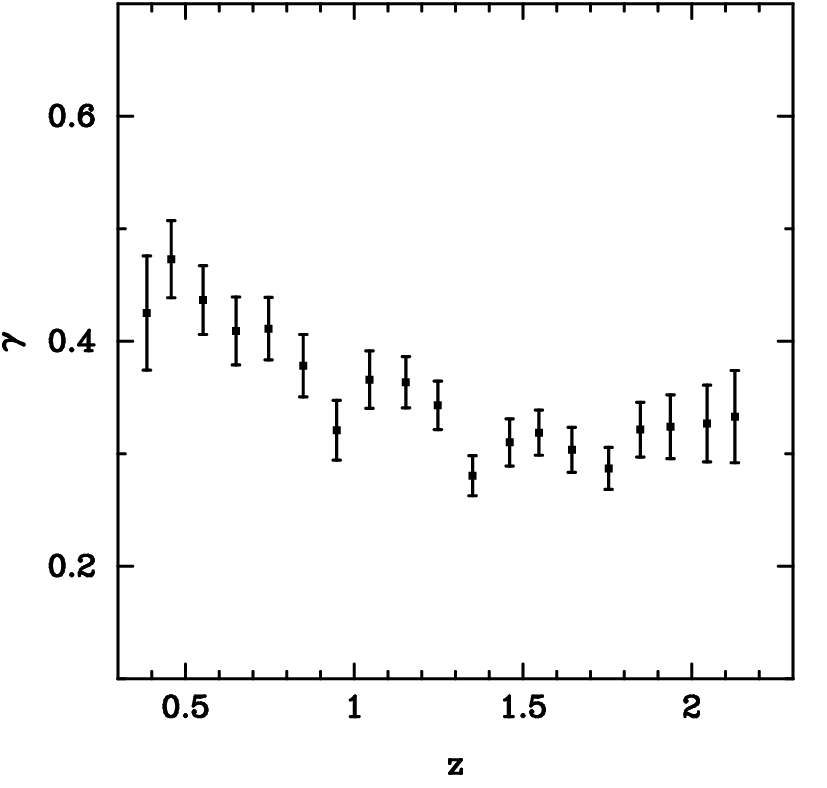}\hfil\plotone{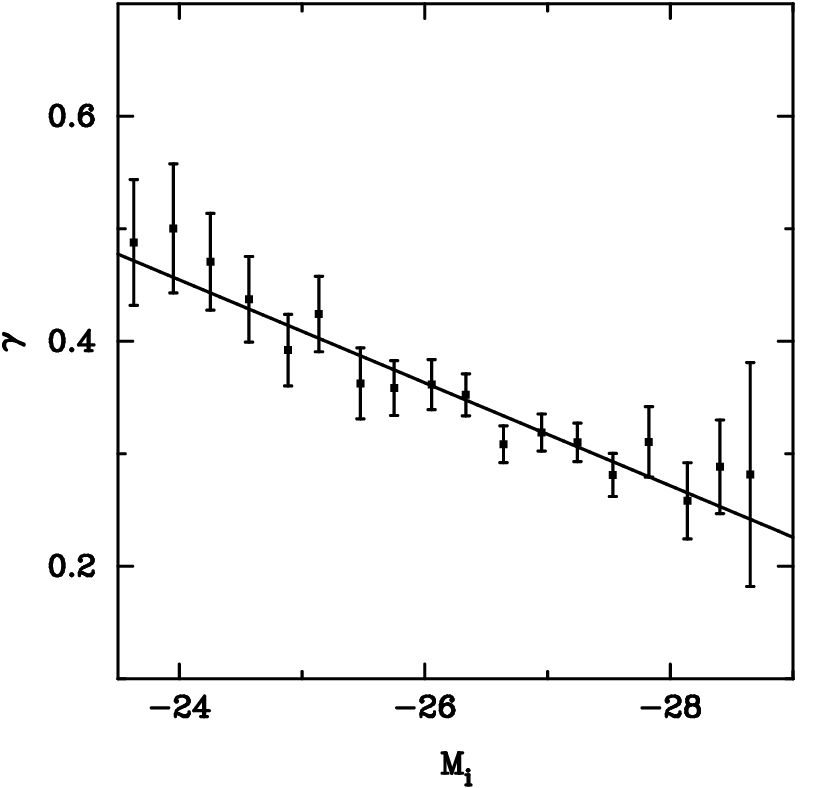}
  \epsscale{1.0}
  \caption{Mean (top) and standard deviation (bottom) of $\ln W$, versus
    redshift (left) or absolute magnitude (right), for \mgii.}
  \label{fig:params-mgii}
\end{figure}

\begin{figure}
  \centering
  \epsscale{0.45}
  \plotone{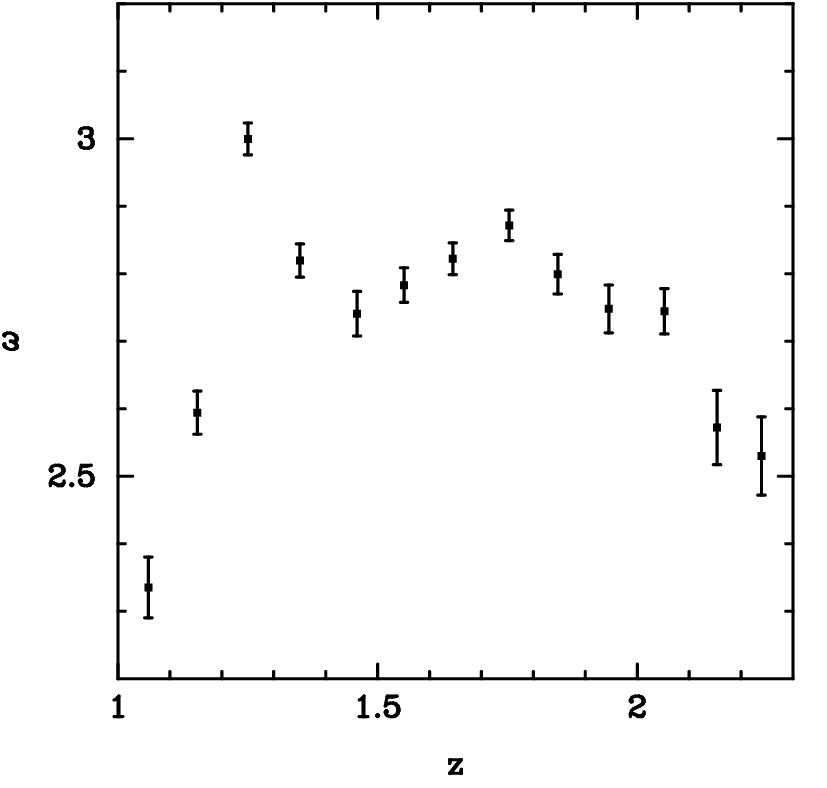}\hfil\plotone{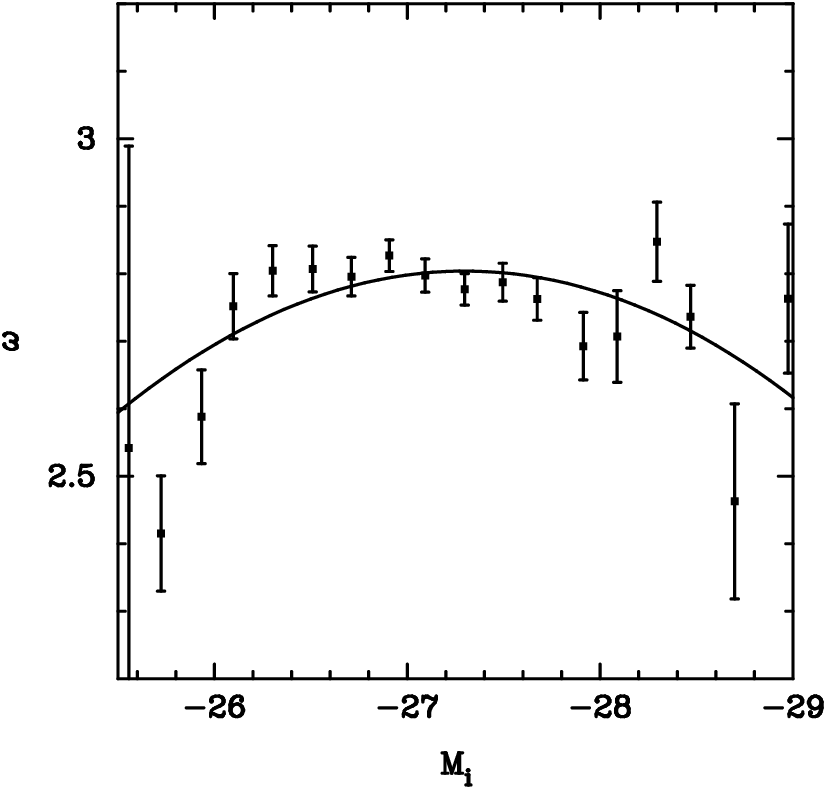}\\
  \plotone{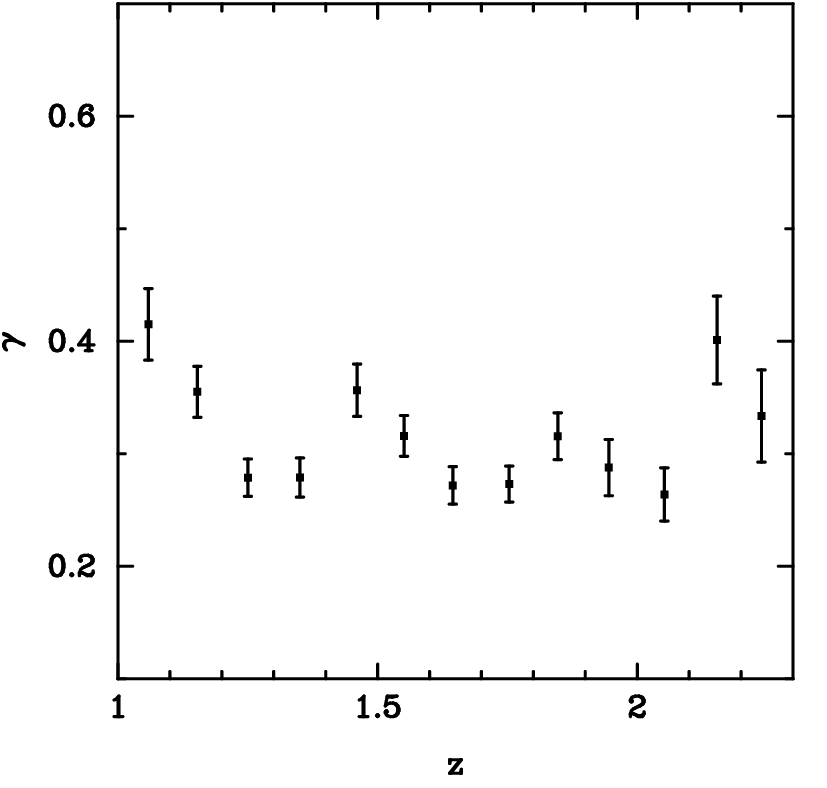}\hfil\plotone{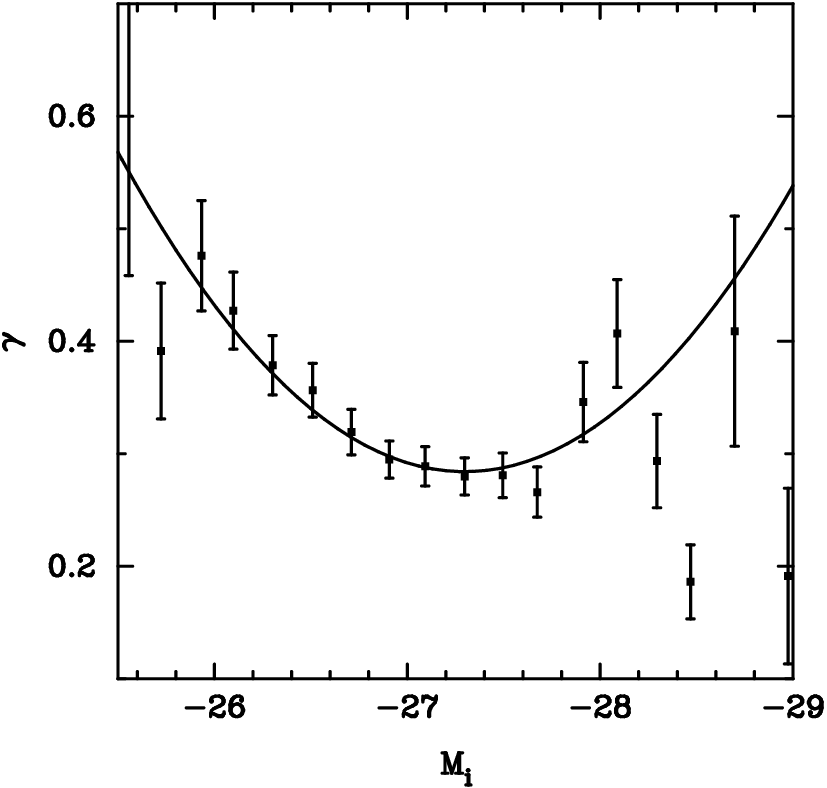}
  \epsscale{1.0}
  \caption{Mean (top) and standard deviation (bottom) of $\ln W$, versus
    redshift (left) or absolute magnitude (right), for \ciii.}
  \label{fig:params-ciii}
\end{figure}

\begin{figure}
  \centering
  \epsscale{0.45}
  \plotone{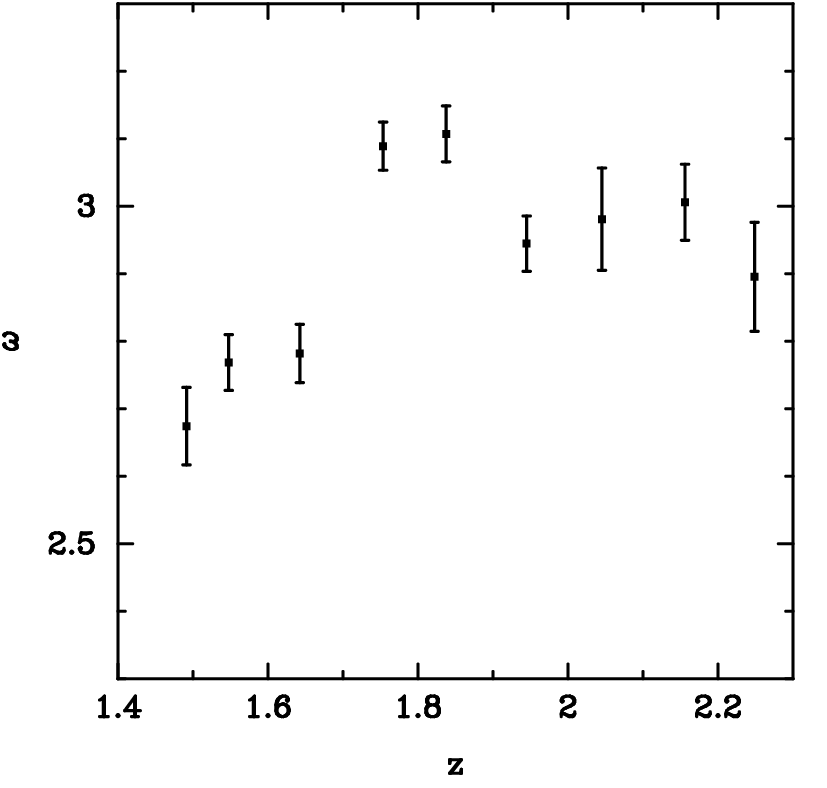}\hfil\plotone{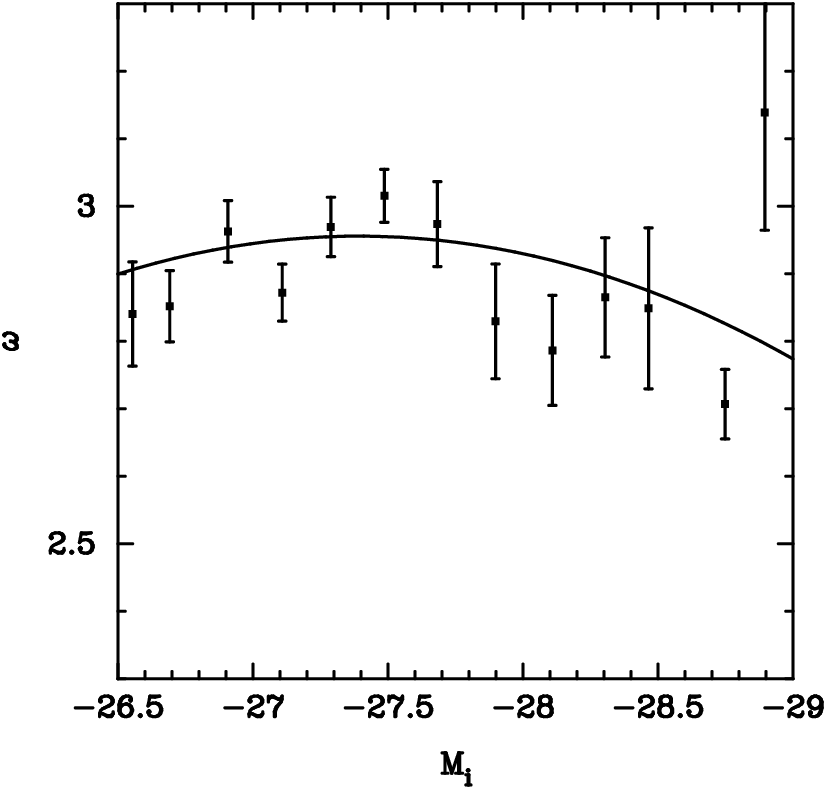}\\
  \plotone{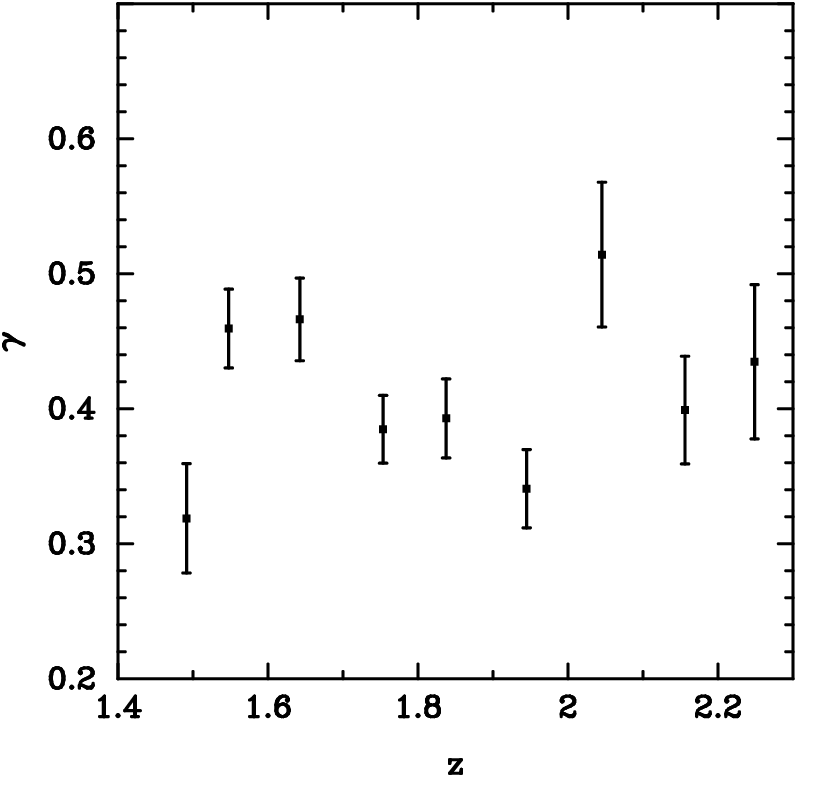}\hfil\plotone{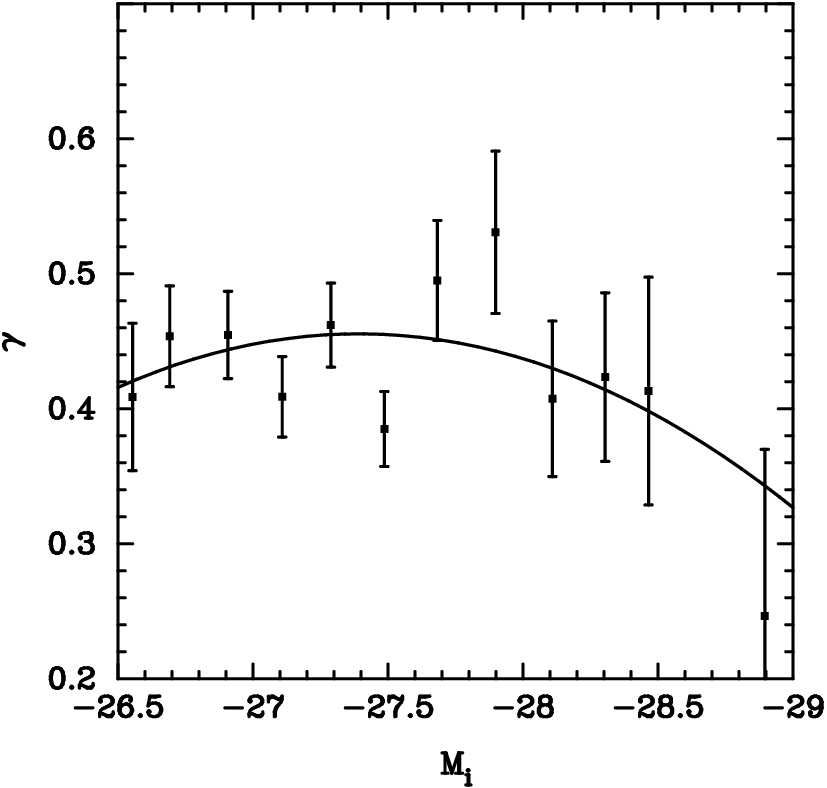}
  \epsscale{1.0}
  \caption{Mean (top) and standard deviation (bottom) of $\ln W$, versus
    redshift (left) or absolute magnitude (right), for \civ.}
  \label{fig:params-civ}
\end{figure}

The purpose of the present work is not to characterize the Baldwin
effect in great detail; however, some basic understanding is needed to
account for this evolution effect in the likelihood function.  To
elucidate whether the $z$ or $M_\iband$ correlation might be stronger, I
again divided the observations into redshift or magnitude bins, and
calculated the mean and standard deviation of $\ln W$ for each bin
(equivalently yielding $\omega$ and $\gamma$ for the best-fit unlensed
lognormal distribution).  The results for each of the three emission
lines are plotted in
Figures~\ref{fig:params-mgii}--\ref{fig:params-civ}.  These plots
suggest that the variation is correlated more strongly with $M_\iband$.

In the absence of an obvious linear relationship for most cases, I
allow the shape parameters to vary with magnitude as follows:
\begin{eqnarray}
  \label{eq:omega-fit}
  \omega_{MgII}(M) & = & \omega_0 + \omega_1(M-M_0) 
  \mbox{ for $M < M_0$ ($\omega_0$ otherwise)} \nonumber\\
  \gamma_{MgII}(M) & = & \gamma_0 + \gamma_1(M-M_0) \nonumber\\
  \omega_{CIII}(M) & = & \omega_0 + \omega_1(M-M_0)^2 \\
  \gamma_{CIII}(M) & = & \gamma_0 + \gamma_1(M-M_0)^2 \nonumber\\
  \omega_{CIV}(M) & = & \omega_0 + \omega_1(M-M_0)^2 \nonumber\\
  \gamma_{CIV}(M) & = & \gamma_0 + \gamma_1(M-M_0)^2 . \nonumber
\end{eqnarray}
The sets of parameters $(M_0, \omega_0, \omega_1, \gamma_0, \gamma_1)$
are of course fit separately for each data set.  There is no real
\emph{a priori} justification for keeping the characteristic magnitude
parameters $M_0$ the same for both the $\omega$ and $\gamma$ fits, other
than to reduce the number of parameters being added to the likelihood
function.

\subsection{Equivalent Width Fractional Error}
\label{sec:ewfracerr}

Each measurement of equivalent width in the quasar catalog $W_i$ is
accompanied by an estimate of the measurement error $\sigma_{W,i}$.  In
principle these errors can be incorporated into the likelihood function;
for example, if the errors are assumed to be normal, then for each data
point Equation~\ref{eq:pW} can be convolved with a Gaussian with a width
equal to the fractional error $\sigma_{W,i}/W_i$.  When there is no
lensing, this is approximately equivalent to adding the fractional error
in quadrature to the lognormal width parameter $\gamma$.  This leads to
slightly smaller determinations of the best-fit $\gamma$, on the order
of $0.02$ for the EDR data, and also tends to shift the best-fit
$\omega$ systematically higher.  The non-zero $\Omega_c$ case is
considerably more difficult to treat in the maximum-likelihood
formulation.  However, for small $\Omega_c$ and typical values of the
lognormal shape parameters, and given that typically $\sigma_{W,i}/W_i
\lesssim 0.2$ for the observations, it can be seen that the convolved
distribution, though slightly wider, does not differ appreciably in the
strength of the low equivalent width lensing signature, so estimates of
$\Omega_c$ will not be significantly affected.  Additionally, since the
intrinsic values of the lognormal shape parameters are not of interest
here, and because accounting for equivalent width measuring error takes
vastly more CPU time, I disregard this contribution.

\subsection{Maximum-Likelihood Analysis}
\label{sec:maxl}

\begin{figure}
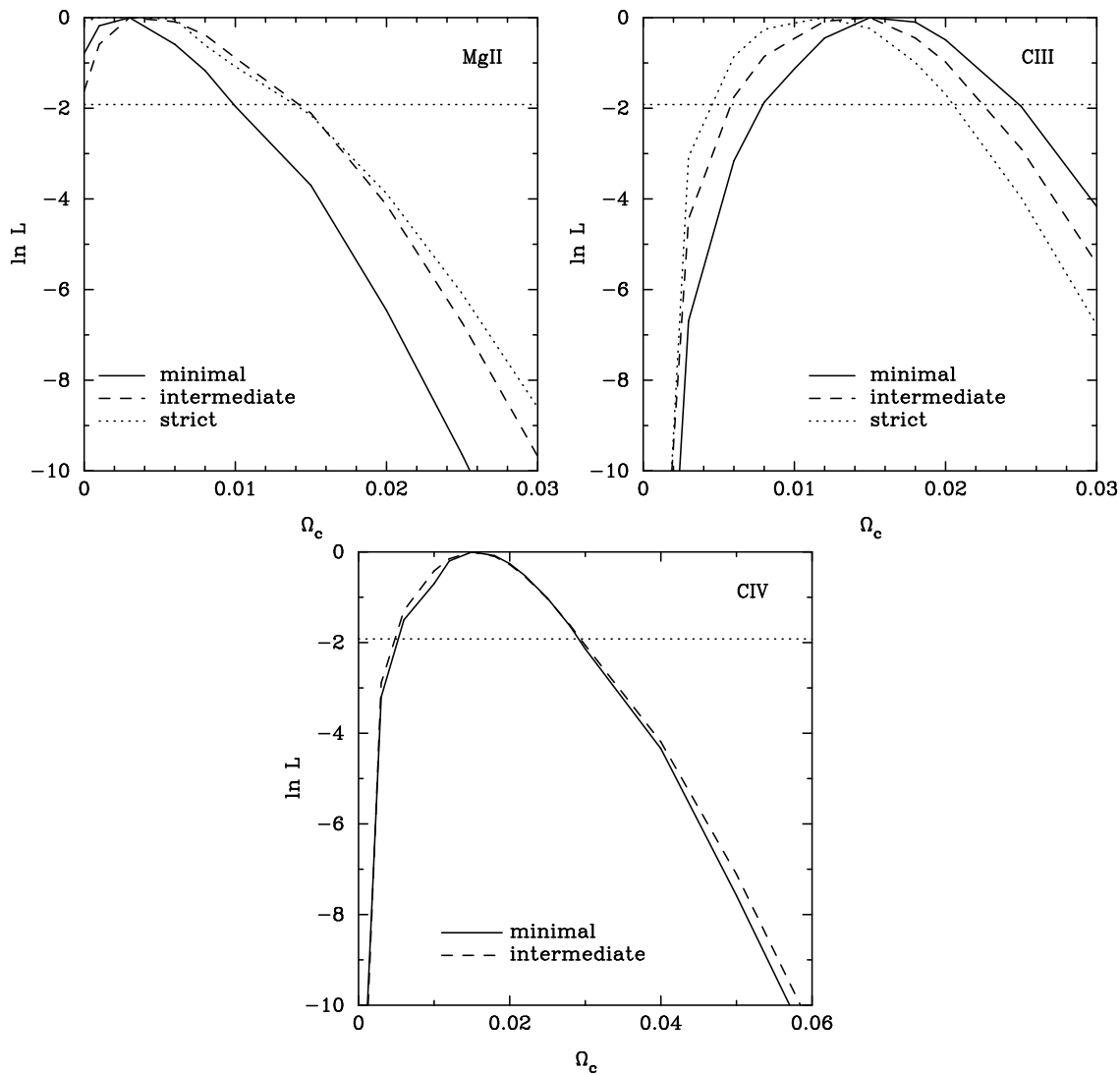

  \epsscale{0.45}
  \plotone{f21a.eps}
  \plotone{f21b.eps}
  \plotone{f21c.eps}
  \epsscale{1.0}
  \caption{Maximized log-likelihood versus $\Omega_c$ for each data set,
    using a model with $M_{lens} = 1\Msolar$.  The lines represent the
    minimal (\emph{solid line}),
    intermediate (\emph{dashed line}),
    and strict (\emph{dotted line}) data quality cuts.}
  \label{fig:maxl-omega-all}
\end{figure}

For each spectral line data set, and using a model in which $M_{lens} =
1\Msolar$ I maximized the log-likelihood function, allowing $\Omega_c$
to be the undetermined parameter and marginalizing over all others.  In
an attempt to determine how the quality of the line fits affects the
results, I made three different ``data quality'' cuts, requiring
($\sigma_W/W$, $\chi^2/\nu$) to be less than (0.5, 5.0), (0.4, 3.0), or
(0.3, 2.0) for ``minimal,'' ``intermediate,'' and ``strict'' cuts.
Results, normalized to a common maximum-likelihood value, are shown in
Figure~\ref{fig:maxl-omega-all}.  The 95\%
confidence level is attained in this one-parameter case when $\ln L =
\ln L_{max} - 1.92$ \citep{book:cowan}.  It is perhaps somewhat
surprising that weakening the data selection criteria do not always
produce more conservative limits for $\Omega_c$.

\begin{figure}
  \plotone{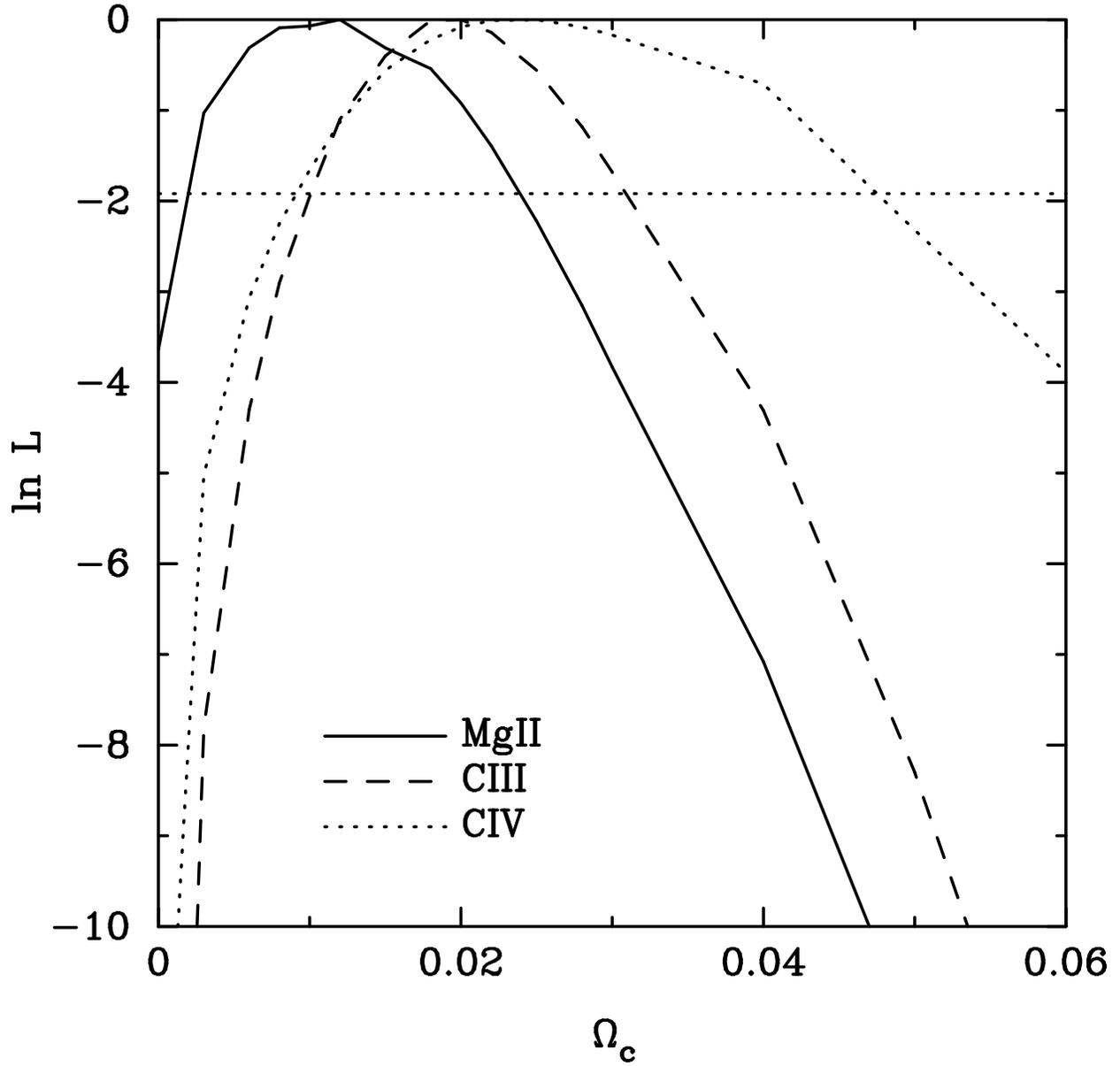}
  \caption{Maximized log-likelihood versus $\Omega_c$ in a
    $M_{lens}=0.001\Msolar$ model, for all three data sets.
    The lines represent \mgii\ (\emph{solid line}), \ciii\ (\emph{dashed
      line}), and \civ\ (\emph{dotted line}) data sets.}
  \label{fig:maxl-omega-all-m001}
\end{figure}

I also consider a $M_{lens}= 0.001\Msolar$ lensing model for each of the
three samples of lines, using the ``intermediate'' selection criteria
for the data.  The maximum-likelihood curves for all three lines are
plotted together in Figure~\ref{fig:maxl-omega-all-m001}.  The limits
from this plot, along with those from the $1\Msolar$ lens model, are
summarized in Table~\ref{tab:omega-limits}.

Goodness of fit can be measured with a $\chi^2$ statistic applied to the
binned data sets.  For the various combinations of the three emission
lines and the two lens masses, $\chi^2/\nu$ values range from 1.2 to
1.6, indicating that the model represents a generally acceptable fit.

\subsection{Tests of Maxmimum-Likelihood Method}
\label{sec:maxl-tests}

Applying the maximum-likelihood analysis to some simple modifications of
the data set serves as an indication of the method's validity and
sensitivity.  For example, one expects that adding or removing low
equivalent width, high redshift observations will increase or decrease
respectively the maximum-likelihood estimator (MLE) of $\Omega_c$ or its
upper bound.  In one such test, I removed the lowest four equivalent
widths (0.4\% of the data set) from the \ciii\ ``intermediate quality''
data set; the widths ranged from 2.3~\AA\ to 4.3~\AA\ and were at
redshifts of 1.1, 1.5, 1.8, and 2.2.  This modification shifted the 95\%
confidence upper limit on $\Omega_c$ from 0.022 to 0.02; while the MLE
value of $\Omega_c$ remained virtually unchanged, the peak was
considerably flattened and the lower bound moved from 0.006 to 0.002.
In a similar test, adding just four counterfeit low equivalent widths
(from 2.0~\AA\ to 2.4~\AA) to the original data set, at moderately high
redshift ($z=1.6$ to $2.0$), was sufficient to shift the upper limit to
0.028 and $\Omega_{c,MLE}$ from 0.015 to 0.018.

I also tested how well the maximum-likelihood method could determine
$\Omega_c$ from synthesized data sets.  Keeping each \ciii\ redshift and
magnitude measurement the same, I replaced the equivalent widths with
Monte Carlo ``data'' generated using the magnitude-dependent lognormal
shape parameters from the actual data set, and assuming $\Omega_c =
0.03$.  The analysis of this fake data appropriately yielded
$\Omega_{c,MLE}=0.032$ with an upper bound of 0.045.  For a lens-free
$\Omega_c=0$ synthetic data set, the maximum-likelihood method correctly
produced a peak at $\Omega_{c,MLE}=0$ with an upper limit of 0.007.

\subsection{Scaling to Larger Samples}
\label{sec:scaling}

\begin{figure}
  \plotone{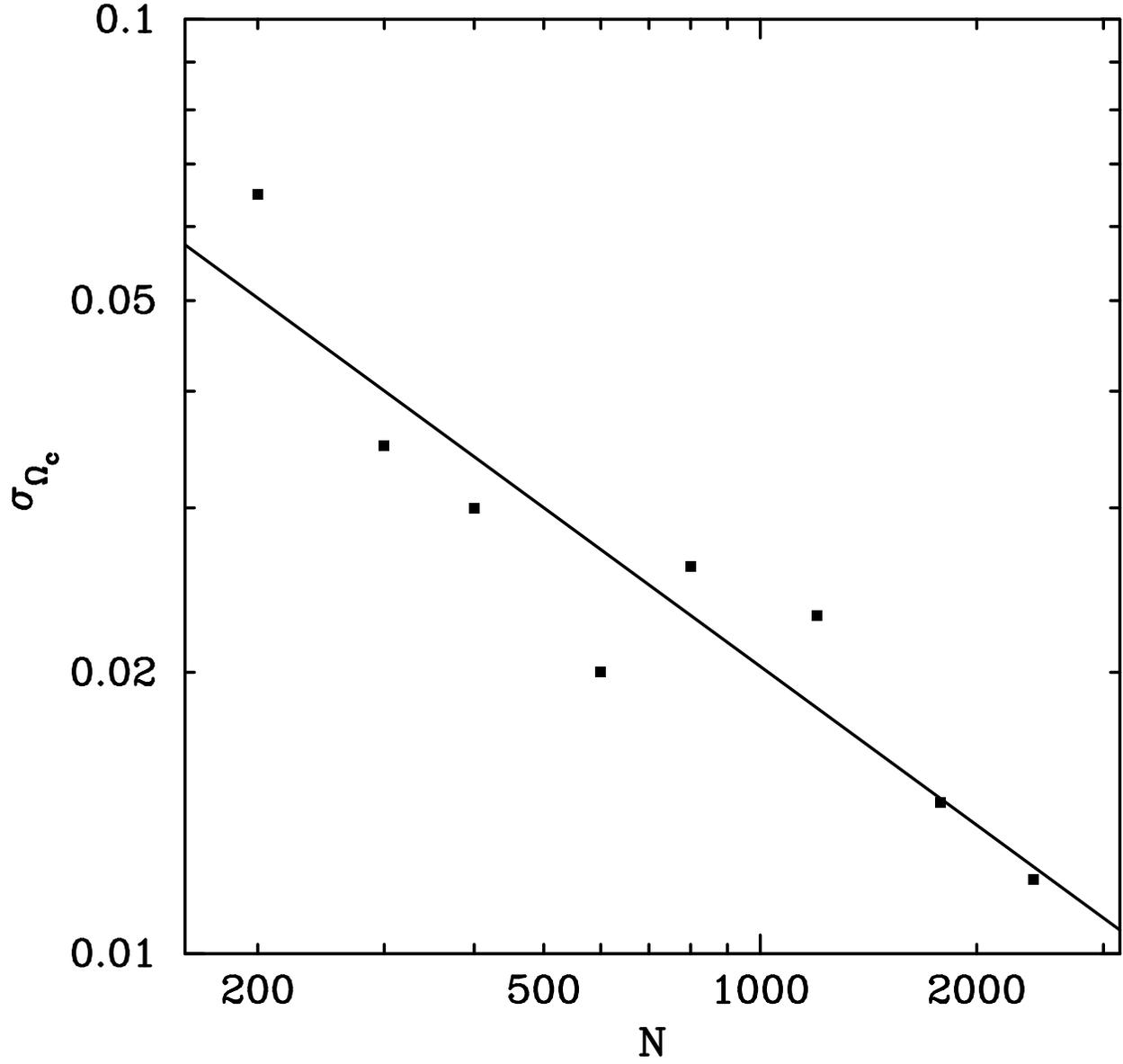}
  \caption{$\sigma_{\Omega_c}$ versus $N$ for Monte Carlo
    samples from the \mgii\ data set.}
  \label{fig:limits-vs-n}
\end{figure}

Anticipating the growth in the SDSS quasar catalog, I have made an
attempt to estimate how the constraints on $\Omega_c$ depend on the
number of equivalent widths measured.  From the \mgii\ data set I
generated random subsamples of various sizes (as well as one supersample
using sampling with replacement), and performed the same
maximum-likelihood analysis in each case.  The results are plotted as
log-limit versus log-number in Figure~\ref{fig:limits-vs-n}.  The slope
of the fit is $-0.56$, indicating that the bound on $\Omega_c$ scales
approximately like $1/\sqrt{N}$.  This makes sense in light of the fact
that the fraction of low equivalent width lines in the lensing model
(i.e., the lensing signal divided by $N$) scales roughly linearly with
$\Omega_c$ for $\Omega_c \ll 1$.  Because the EDR Quasar Catalog
represents about 4\% of the SDSS goal for quasar observations, this
implies it may be possible to improve the constraints by a factor of 5.

\section{Conclusions}
\label{cha:conclude}

I have shown that microlensing by cosmologically distributed compact
objects in the mass range $0.001$--$1\Msolar$ can produce a detectable
increase in the fraction of small equivalent width quasar emission lines
with redshift.  Table~\ref{tab:omega-limits} details the limits derived
from each emission line studied and assuming lens masses at either
extreme of the range.  To summarize these results, and based on the lack
of a strong lensing signal, $\Omega_c$ is constrained to be less than
approximately $0.03$ at 95\% confidence over this mass range.

\begin{deluxetable}{ccc}
  \tablewidth{0pt}
  \tablecaption{Upper limits (95\% confidence level) on cosmological
    compact dark matter.\label{tab:omega-limits}}
  \tablehead{\colhead{Spectral Line} & 
    \colhead{$M_{lens}=1\Msolar$} & 
    \colhead{$M_{lens}=0.001\Msolar$}}
  \startdata
  \mgii & $\Omega_c < 0.014 $ & $\Omega_c < 0.024 $ \\
  \ciii & $\Omega_c < 0.025 $ & $\Omega_c < 0.031 $ \\
  \civ  & $\Omega_c < 0.029 $ & $\Omega_c < 0.047 $ \\
  \enddata
\end{deluxetable}

This limit is interesting in several respects.  First of all, it is
stricter than limits derived from the current microlensing searches
within our own halo (e.g., \citealt{lens:macho-lmc2000}); if at most
one-third of the Galactic halo mass is compact, then their results imply
$\Omega_c \lesssim 0.1$ for $\Omega_M = 0.3$.  Additionally, since
$\Omega_c < \Omega_B \approx 0.04$, it may not ultimately be necessary
to invoke exotic non-baryonic compact objects to explain microlensing
events.  This conclusion will be made much stronger if larger future
quasar samples demonstrate the limit on $\Omega_c$ to be $\lesssim
0.01$, the fraction of baryons unexplained in the current census
\citep{misc:silk-baryons}.  Also below this threshold, microlensing by
ordinary stars in galaxies may become more significant, since
$\Omega_\ast \approx 0.006$ \citep*{misc:baryon-budget}.

There are some important points to keep in mind when considering the
analysis leading to these constraints.  The first is that, because the
three sets of emission lines are from only one sample of quasars, the
constraints are \emph{not} independent and cannot in general be combined
to derive a stronger limit.  Rather the three sets of limits serve in
some sense as checks on one another.  The second point concerns the
absence of quoted lower limits.  While the maximum-likelihood analysis
formally produces a nonzero most-likely value for $\Omega_c$ in some
cases, along with a lower limit, this should not be interpreted as an
unambiguous detection of compact dark matter.  It may very well be that
the intrinsic distribution at low equivalent widths departs from a
simple lognormal; this could be easily be misconstrued as evidence for
lensing at low $\Omega_c$.  Another way of saying this is that because
probability must be non-negative, any departure from the exponentially
small intrinsic probability at low equivalent width will result in a
positive detection of compact dark matter, real or not.  The other side
of this argument is that if the actual unlensed distribution does in
fact have some probability at low equivalent widths not represented in
the model, then the limits from the simple lognormal-based model will
necessarily be conservative.

A large source of systematic uncertainty in this analysis is the Baldwin
effect, which deserves a much more thorough treatment than was possible
in this paper.  The maximum-likelihood requirements of an adequately
parameterized distribution has led to the introduction of a number of
\emph{ad hoc} parameters irrelevant to the final result.  Much work
remains to be done in characterizing the luminosity evolution of
equivalent width.  It would also be worthwhile to consider alternate
methods of constructing statistics for determining $\Omega_c$ that would
rely less on knowing the details of the Baldwin effect while still using
the sheer numbers of observations to full advantage.

An additional source of systematic error is the single-component
Gaussian line-fitting process in the SDSS spectroscopic pipeline.
Quasar broad emission lines are often fit more closely by using an
additional Gaussian to incorporate the flux in the broad low wings.
Even this can fail to account fully for the wavelength shifts and line
asymmetries often found in quasar spectra \citep{sdss:qso-belr-shifts}.
Some recent work within the SDSS Collaboration has focused on improving
the current quasar line fitting algorithm \citetext{D. Vanden Berk 2003,
  private communication}.

The lensing model is another area in which much improvement can be made.
The assumption of uniformly distributed cosmological lenses in the
spirit of \citet{lens:press-gunn}, although analytically simple, is
violated by the clustering actually observed in the universe.  While
\citet{lens:dalcanton} convincingly argue that compact object
correlations are unimportant for the relatively small lensing optical
depths under consideration, \citet{lens:wyithe2002} suggest that lens
clustering in galactic dark matter halos might decrease the fraction of
microlensed sources (and hence weakening the constraints on $\Omega_c$)
by as much as a factor of two.  Moreover, the overall magnification
distribution can be altered by considering the effects of smooth dark
matter in halos, in which the compact lens may reside.
\citet{lens:extragal-machos} examine the possibility of using this
quasar equivalent width method to detect the signature of extragalactic
halo MACHOs in the final SDSS quasar catalog.
\citet*{lens:mortsell2001} and \citet{lens:metcalf1999} consider models
with both smooth and compact dark matter components in the context of
lensed Type~Ia supernovae.  \citet{lens:seljak1999} have also made
predictions for SNe~Ia lensing by incorporating large scale structure
information from N-body simulations.

Finally, the data set itself will only become more powerful as the SDSS
continues to increase its tally of spectroscopically observed quasars.
When the SDSS quasar luminosity function is determined, it will remove
the mismatch between $B$ and $i$ magnitudes that is a concern with the
current use of the 2dF luminosity function.  The large quasar sample
will also permit a much more detailed characterization of emission line
properties that will hopefully lead to a better understanding of quasar
structure, and hence of the effects of quasar microlensing.

\acknowledgments

Thanks go to my advisor, Josh Frieman, for many helpful discussions
underlying this project, and for his willingness and ability to advise
on a long-distance basis.  Mark SubbaRao lent his expertise with both
the SDSS catalog database and the spectroscopic pipeline.  Julianne
Dalcanton furnished valuable initial insight into the applications of
the lensing model to data sets.  I also gratefully acknowledge the
support provided by Department of Energy grant DE-FG02-90ER40560.

Funding for the creation and distribution of the SDSS Archive has been
provided by the Alfred P. Sloan Foundation, the Participating
Institutions, the National Aeronautics and Space Administration, the
National Science Foundation, the U.S. Department of Energy, the Japanese
Monbukagakusho, and the Max Planck Society. The SDSS Web site is
\url{http://www.sdss.org/}.

The SDSS is managed by the Astrophysical Research Consortium (ARC) for
the Participating Institutions. The Participating Institutions are The
University of Chicago, Fermilab, the Institute for Advanced Study, the
Japan Participation Group, The Johns Hopkins University, Los Alamos
National Laboratory, the Max-Planck-Institute for Astronomy (MPIA), the
Max-Planck-Institute for Astrophysics (MPA), New Mexico State
University, University of Pittsburgh, Princeton University, the United
States Naval Observatory, and the University of Washington.

\bibliographystyle{apj}
\bibliography{apj-jour,sdss,2dF,lens,qso,baldwin,book,misc}

\end{document}